\documentclass[prl,cn,twocolumn,superscriptaddress]{revtex4}

\usepackage{amsmath}
\usepackage{amssymb}
\usepackage{amsthm}
\usepackage{amsfonts}
\usepackage{listings}
\usepackage{enumerate}
\usepackage{latexsym}
\usepackage{color}
\usepackage{bm}
\usepackage{hyperref}
\hypersetup{
 pdfnewwindow=true, colorlinks=true,
 linkcolor=blue, anchorcolor=blue,
 citecolor=blue, filecolor=blue,
 menucolor=blue, urlcolor=blue}

\newcommand{\beginsupplement}{%
        \setcounter{table}{0}
        \renewcommand{\thetable}{S\arabic{table}}%
        \setcounter{figure}{0}
        \renewcommand{\thefigure}{S\arabic{figure}}%
     }

\usepackage{psfrag}

\usepackage{bm}
\usepackage{graphicx}
\usepackage{subfigure}

\RequirePackage[normalem]{ulem} 
\RequirePackage{color}\definecolor{RED}{rgb}{1,0,0}\definecolor{BLUE}{rgb}{0,0,1} 

\newcommand{\braket}[2]{\left\langle #1 | #2 \right\rangle}

\newcommand{\ket}[1]{\left|#1\right\rangle}

\newcommand{\up}{\uparrow}
\newcommand{\dw}{\downarrow}
\newcommand{\bb}{{\bf b}}
\newcommand{\bk}{{\bf k}}

\def \tic{Cu$_3$SbS$_4$}
\def \wsmc{Cu$_2$ZnGeSe$_4$}
\def \ti{Case I}
\def \wsm{Case II}
\def\ie{{\it i.e.},\ }
\def\eg{{\it e.g.}\ }

\input{epsf}

\begin{document}

\tolerance 10000

\newcommand{\vk}{{\bf k}}

\draft

\title{Weyl Semimetals with $S_4$ symmetry}


\author{Yuting Qian}
\thanks{These authors contributed equally to this work.}
\affiliation{Beijing National Laboratory for Condensed Matter Physics,
and Institute of Physics, Chinese Academy of Sciences, Beijing 100190, China}
\affiliation{University of Chinese Academy of Sciences, Beijing 100049, China}

\author{Jiacheng Gao}
\thanks{These authors contributed equally to this work.}
\affiliation{Beijing National Laboratory for Condensed Matter Physics,
and Institute of Physics, Chinese Academy of Sciences, Beijing 100190, China}
\affiliation{University of Chinese Academy of Sciences, Beijing 100049, China}

\author{Zhida Song}
\affiliation{Department of Physics, Princeton University, Princeton, NJ 08544}

\author{Simin Nie}
\affiliation{Department of Materials Science and Engineering, Stanford University, Stanford, California 94305, USA}

\author{Zhijun Wang}
\email{wzj@iphy.ac.cn}
\affiliation{Beijing National Laboratory for Condensed Matter Physics,
and Institute of Physics, Chinese Academy of Sciences, Beijing 100190, China}
\affiliation{University of Chinese Academy of Sciences, Beijing 100049, China}

\author{Hongming Weng}
\email{hmweng@iphy.ac.cn}
\affiliation{Beijing National Laboratory for Condensed Matter Physics,
and Institute of Physics, Chinese Academy of Sciences, Beijing 100190, China}
\affiliation{University of Chinese Academy of Sciences, Beijing 100049, China}
\affiliation{Songshan Lake Materials Laboratory, Dongguan, Guangdong 523808, China}
\affiliation{CAS Center for Excellence in Topological Quantum Computation, University of Chinese Academy of Sciences, Beijing 100190, China}
\author{Zhong Fang}
\affiliation{Beijing National Laboratory for Condensed Matter Physics,
and Institute of Physics, Chinese Academy of Sciences, Beijing 100190, China}
\affiliation{University of Chinese Academy of Sciences, Beijing 100049, China}

\begin{abstract}
In the time-reversal-breaking centrosymmetric systems, the appearance of Weyl points can be guaranteed by an odd number of all the even/odd parity occupied bands at eight inversion-symmetry-invariant momenta. Here, based on symmetry analysis and first-principles calculations, we demonstrate that for the time-reversal-invariant systems with $S_4$ symmetry, the Weyl semimetal phase can be characterized by the inequality between a well-defined invariant $\eta$ and an $S_4$ indicator $z_2$. By applying this criterion, we find that some candidates, previously predicted to be topological insulators, are actually Weyl semimetals in the noncentrosymmetric space group with $S_4$ symmetry. Our first-principles calculations show that four pairs of Weyl points are located in the $k_{x,y}$ = 0 planes, with each plane containing four same-chirality Weyl points. An effective model has been built and captures the nontrivial topology in these materials. Our strategy to find the Weyl points by using symmetry indicators and invariants opens a new route to search for Weyl semimetals in the time-reversal-invariant systems.
\end{abstract}

\maketitle
\section{Introduction}
\begin{figure}[!tb]
\centering
\includegraphics[width=7.2 cm]{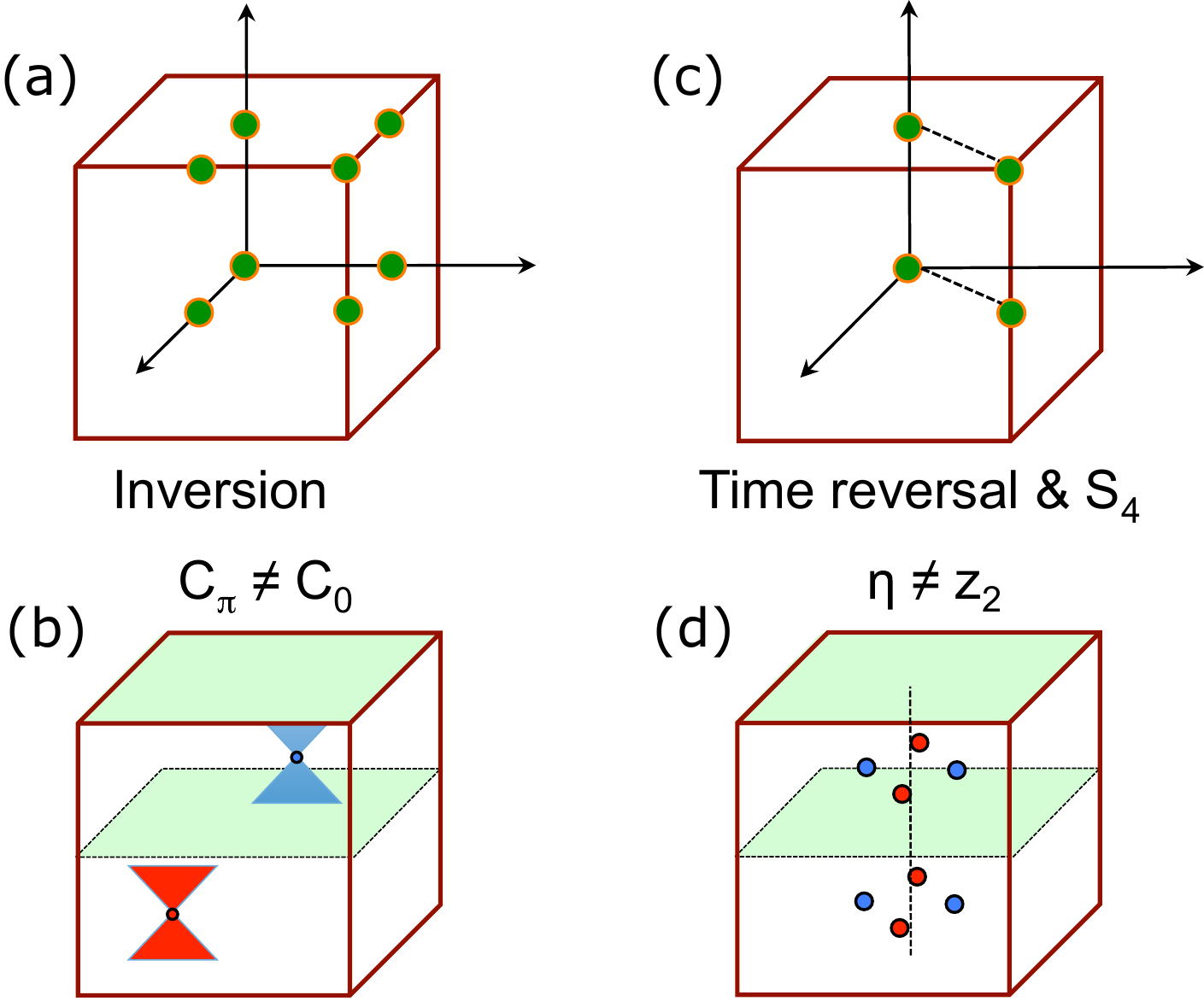}
\caption{(Color online)
Schematic WSMs with symmetry indicators and topological invariants.
For a TRB centrosymmetric system, an odd number of all the even/odd parity occupied bands at eight ISI momenta [green dots in ($\bold a$)] reveals that the Chern numbers of the two ISI planes are different, which guarantee the appearance of odd pairs of Weyl points in 3D Brillouin zone (BZ) ($\bold b$). Note that there is always an even number of even/odd parity occupied bands in a TRI system.
For a TRI and $S_4$-symmetric system, a $z_2$ indicator is defined on four $S_4$ invariant momenta [green dots in ($\bold c$)], and the inequality between the invariant $\eta$ [defined in the main text] and $S_4~z_2$ indicator reveals the appearance of Weyl points, as shown in ($\bold d$). The red (blue) dots stand for +1 (-1) chiral Weyl points.
} \label{fig:1}
\end{figure}

Topological materials~\cite{RevModPhys.90.015001,Wan2011,Weng2016Topological,Wang2013,Qi2010The,Hasan2010Topological,Bernevig2006Quantum,Zhang2009Topological,PRBwzj,Wang2016} have attracted a lot of attentions in the past decades. Many candidates of topological insulators (TIs) are predicted theoretically first, and verified experimentally later~\cite{Bernevig2006Quantum,konig2007quantum,Zhang2009Topological,chen2009experimental,advwang}. Most of the predictions are indicated by topological invariants or symmetry indicators~\cite{PhysRevLett.98.106803,Fu2007IS,Haruki2018,song2017,nature_2019,Zhang2018,wanxg2019,Slager2017}. 
Topological Weyl semimetals  (WSMs)\cite{Murakami_2007,Liu2014Weyl,wengprx,Soluyanov2015Type,Weng2016Coexistence,Lv2015Observation,Lv2015Experimental,nie2017topological,PhysRevLett.117.236401} show linear dispersion around discrete doubly-degenerate points, termed the Weyl points, which are regarded as the sinks/sources of Berry curvature in momentum space. They exhibit many exotic properties, such as Fermi-arc states~\cite{xu2016observation,xu2015discovery,wang2016observation2} on the surfaces, chiral anomaly~\cite{huang2015observation,zhang2016signatures}, and anomalous Hall effect~\cite{Xu2011Chern,Fang92}, etc. However, as Weyl points in the three-dimension (3D) momentum space do {\emph not} require any specific symmetry protection (but the lattice translation symmetry), WSMs~usually can \emph{not} be predicted based on topological invariants or symmetry indicators in the time-reversal-invariant (TRI) systems. As we know, for the time-reversal-breaking (TRB) centrosymmetric systems, the appearance of Weyl points can be guaranteed by an odd number of all the even/odd parity occupied bands at eight inversion-symmetry-invariant (ISI) momenta~\cite{PhysRevLett.117.236401,Hughes2010Inversion,nie2019magnetic}, which can be simply understood by two unequal Chern numbers (if well-defined) of two parallel ISI planes [shown in Figs.~\ref{fig:1}(a) and \ref{fig:1}(b)]. Here, our aim is to find proper topological invariants or symmetry indicators in the TRI systems, which warrant the WSM phase.

In this work, we focus on the TRI systems with $S_4$ symmetry (a more general classification is present in Ref.~\cite{tobedone2019}).
For these systems, we define a topological invariant $\eta$ as
\begin{equation*}
(-1)^{\eta}=(-1)^{\nu_{a_1}}(-1)^{\nu_{a_2}},
\end{equation*}
 which is well defined as long as there are two gapped parallel TRI planes (\eg $a_1$-plane and $a_2$-plane).
The invariants, $\nu_{a_1}$ and $\nu_{a_2}$, are the time-reversal $\mathbb Z_2$ invariants~\cite{Kane2005} of the two parallel TRI planes, respectively.
In addition, the $S_4$ symmetry defines a symmetry indicator $z_2$~\cite{song2017,Haruki2018}.
 Note that a centrosymmetric TRI system always satisfies $\eta=z_2$ if they are well defined~\cite{PhysRevLett.98.106803,Haruki2018}. Here, we find that the inequality between $\eta$ and $z_2$ indicates the appearance of Weyl points generally (without considering additional symmetries). Explicitly, a candidate with $\eta\neq z_2$ can be a WSM, as shown in Figs.~\ref{fig:1}(b) and~\ref{fig:1}(d).

Several years ago, many compounds were predicted to be TIs in the noncentrosymmetric structure of space group 121 ($I\bar 42m$)~\cite{Wang2010}.  However, after we have carefully investigated these so-called ``TIs", we find that they can actually be classified into two different cases based on the $S_4$ indicator: $z_2=1$ (\ti) and $z_2=0$ (\wsm).
In this work, we demonstrate that the ``TIs" in \wsm~virtually turn out to be WSMs. Four pairs of Weyl points are found in the $k_{x,y}=0$ planes, with each plane containing four same-chirality Weyl points. Moreover, the Weyl points are located exactly at the charge neutrality level. The WSM phase is characterized by the inequality between $\eta$ and $z_2$  (\ie $\eta\neq z_2$), which is also applicable to
the WSMs in other space groups with $S_4$ symmetry~\cite{Haijun2016,HgTenc2016}. To capture the nontrivial topology of the materials in space group 121, we have constructed a six-band low-energy effective model. Fermi arcs as iconic surface states of the WSM have also been presented.
Our strategy to find the Weyl points by using symmetry indicators and invariants opens a new route to search for Weyl semimetals in the TRI systems.



\section{Crystal structure and Methodology}
\begin{figure}[!tb]
\centering
\includegraphics[width=8.5 cm]{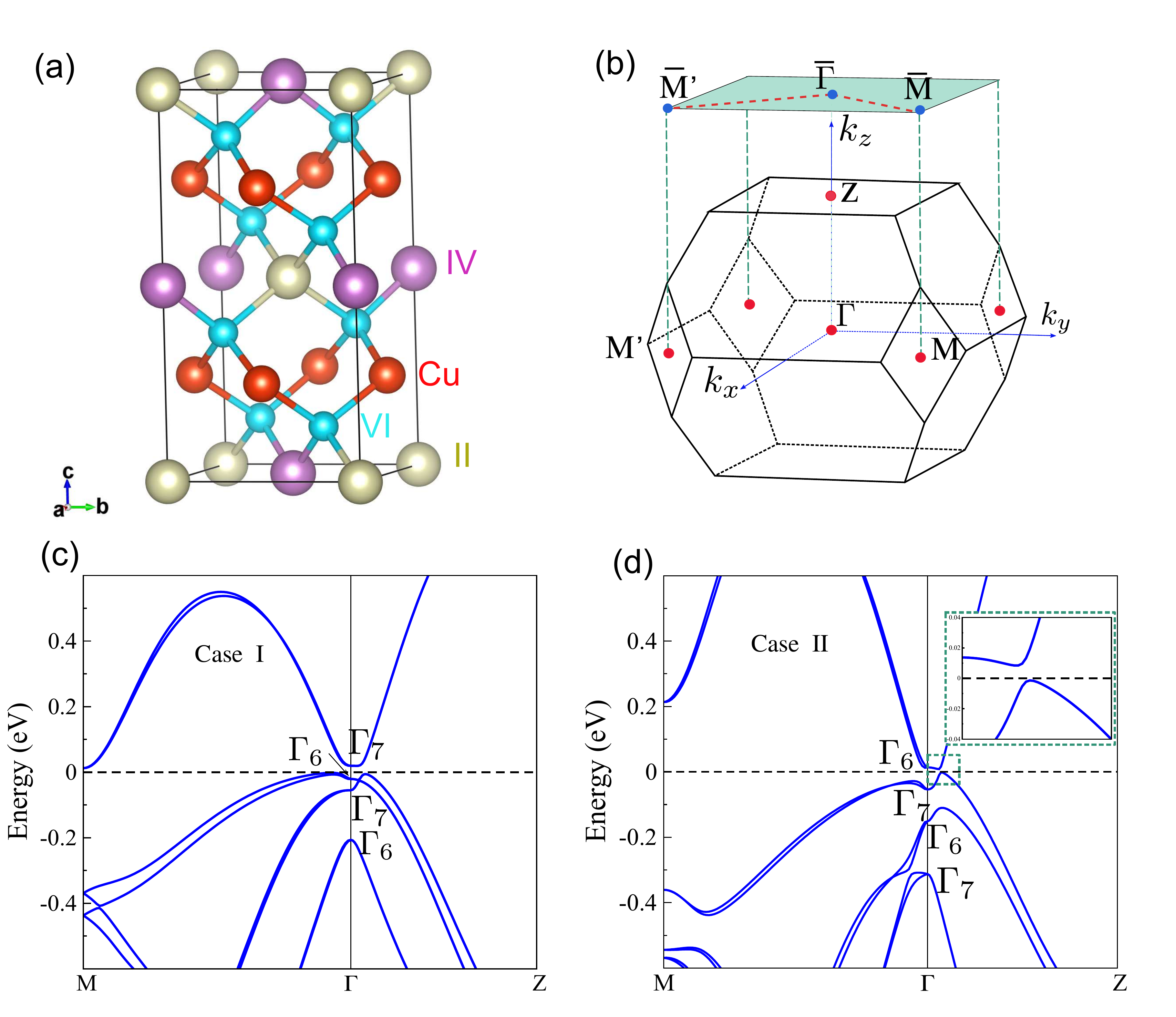}
\caption{(Color online)
(a) Crystal structure of the quaternary stannite Cu$_2$-II-IV-VI$_4$ compounds and (b) BZ for the series of compounds in space group 121. There are alternating cation layers of mixed II and IV atoms, which are separated from each other by layers of Cu monovalent cations. Two equivalent Cu atoms, one II atom, one IV atom and four VI atoms occupy the $4d$, $2a$, $2b$ and $8i$ Wyckoff positions, respectively. In the Cu$_2$-Cu-Sb-VI$_4$ structure, the $2a$ and $2b$ positions are occupied by Cu and Sb atoms, respectively.
The electronic band structures and irreps at $\Gamma$ point with SOC for (c) \tic~and (d) \wsmc~are presented for \ti~ and \wsm, respectively.
} \label{fig:2}
\end{figure}

We investigated a series of Cu-based chalcogenides in the stannite structure: Cu$_2$-Cu-Sb-VI$_4$ and Cu$_2$-II-IV-VI$_4$ with II=$\{$Cd, Hg, and Zn$\}$, IV=$\{$Si, Ge, and Sn$\}$, and VI=$\{$S, Se, and Te$\}$. The series of compounds in space group $I \bar 4 2 m$ ($D_{2d} $) have a body-centered tetragonal crystal structure with lattice parameters: $a$ and $c$.
The structure has three twofold rotational symmetries ($C_{2x,2y,2z}$), two mirror symmetries ($M_{xy,\bar xy}$), and the combined S$_4$ symmetry of inversion symmetry ($I$) and the fourfold rotation ($C_{4z}$). But, neither $I$ nor $C_{4z}$ is respected.
Fig.~\ref{fig:2}(a) presents the stannite structure.
Each anion is tetrahedrally coordinated by four cations with three inequivalent bonds: VI-Cu, VI-II, and VI-IV.
The crystal structure is nearly double zinc-blende structure along $c$ axis but with a little distortion characterized by $c\neq2a$, due to the interlayer coupling. These compounds represent the strained HgTe-class materials~\cite{HgTenc2016}.

We performed the first-principles calculations with VASP package \cite{KRESSE199615,vasp} based on the density functional theory (DFT) with the projector augmented wave (PAW) method \cite{paw1,paw2}.
 The generalized gradient approximation (GGA) with exchange-correlation functional of Perdew, Burke and Ernzerhof (PBE) for  the exchange-correlation functional \cite{pbe} was employed.
The kinetic energy cutoff  was set to 400 eV for the plane wave basis.
A 10$\times$10 $\times$10 $k$-mesh in the self-consistent process for the BZ sampling was adopted. The lattice and atomic parameters in the Inorganic Crystal Structure Database (ICSD) were employed in our calculations, as shown in Table~\ref{tab:s1} in Section A of the Supplemental Material (SM A). The electronic structures with spin-orbit coupling (SOC) were carried out. The Wilson-loop technique~\cite{Yu2011An} was used to calculate topological invariants and chiral charges~\cite{Fang92,Balents2011Weyl}.

\begin{figure*}[!htb]
\centering
\includegraphics[width=16.5 cm]{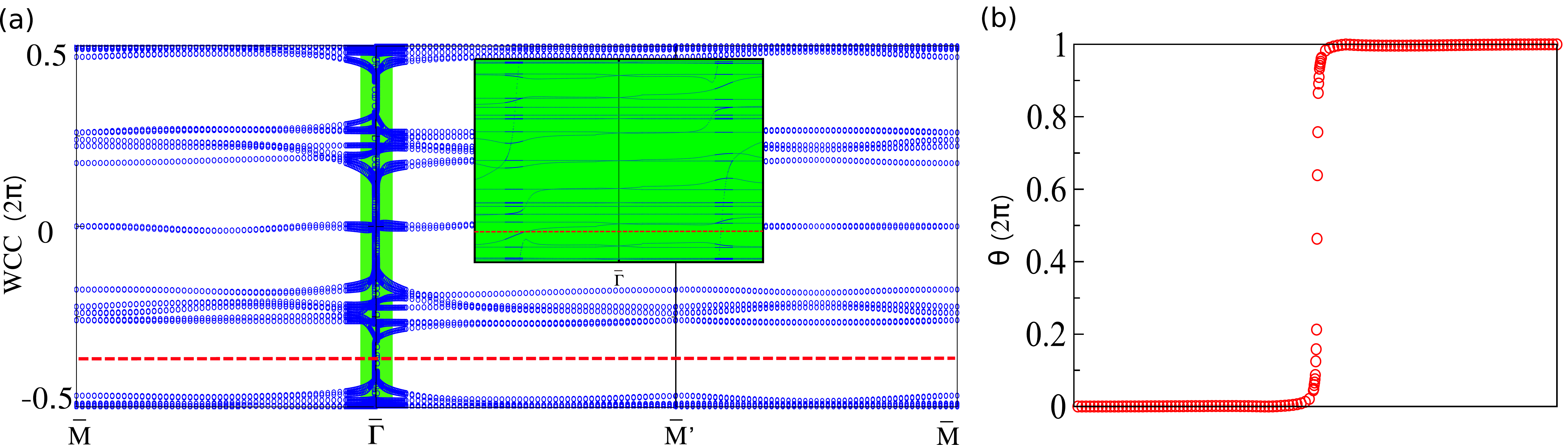}
\caption{(Color online)
(a) The WCC of $k_z$-directed Wilson loops of \wsmc~on the path $\bar M (0.5,0.5)-\bar \Gamma(0,0)-\bar M'(0.5,-0.5)-\bar M(0.5,0.5)$ as marked in Fig.~\ref{fig:2}. (b) The Chirality of the Weyl point at ($0.0036, 0.0, 0.0657$) calculated by Wilson-loop method on a manifold enclosing it.
}
\label{fig:3}
\end{figure*}

\section{Results and discussions}

\subsection{Electronic band structures}

Based on the first-principles calculations, we have reinvestigated the electronic band structures of the compounds, which are proposed to be TIs in the previous work~\cite{Wang2010}.
We find that these compounds can be actually classified into two cases.
In the main text, we take \tic~(\ti) and \wsmc~(\wsm) compounds as two examples of the two cases, respectively, and present the results of the other candidates in the SM.
The calculated band structures along high-symmetry lines are presented in Fig.~\ref{fig:2}.
One can find that there is a band gap along the high-symmetry lines for both compounds.
Then, we have calculated the time-reversal $\mathbb Z_2$ invariants in both $k_z=0$ and $k_z=\frac{\pi}{c}$ planes.
The results of the Wilson loop calculations are presented in Fig.~\ref{fig:s2} of the SM A.
The two $\mathbb Z_2$ invariants are computed to be $\nu_{k_z=0}=1$ and $\nu_{k_z=\frac{\pi}{c}}=0$, giving rise to $\eta=1$ (or $\nu_0=1$ if the system is fully gapped in the 3D BZ~\cite{PhysRevLett.98.106803}). These results seem to be consistent with the previous prediction of ``TIs" \cite{Wang2010}. In this letter, the ``TIs" refer to these topological compounds previously predicted in space group 121, which host $\eta=1$.

Then, we have further checked the irreducible representations (irreps) of the crystal symmetry. In Case I, the $\Gamma_7$ band is the lowest conduction band (LCB) at the $\Gamma$ point, while the LCB is the $\Gamma_6$ band in \wsm.
The $\Gamma_6$ and $\Gamma_7$ irreps are labeled by the little group at $\Gamma$ (the double group of $D_{2d}$).
In a body-centered structure, four $S_{4z}$ invariant momenta (SIM) are $\Gamma[0,0,0]$, $C[0,0,1]$, $A[0.5,0.5,0.5]$, and $B[0.5,0.5,-0.5]$ (hereafter, all $k$ points are given in units of [$\frac{2\pi}{a},\frac{2\pi}{a},\frac{2\pi}{c}$] in Cartesian coordinates).
Note that the $A$ and $B$ points are not TRI momenta. Since $S_4^4=-1$ in a spinful system, the eigenvalues of $S_4$ symmetry are given as $\lambda_j=e^{i\pi\frac{2j-1}{4}}$ with $j\in\{0,1,2,3\}$. The $z_2$ indicator of $S_4$ symmetry in a body-centered structure is defined explicitly as below:
\begin{equation*}
  z_2 = \sum_{K\in\{\Gamma,C,A,B\}}\frac{n_K^{2} - n_K^{0}}{2} \quad{\rm mod} \ 2 ,
\end{equation*}
with $n_K^{i}$ the number of the occupied bands with $S_4$ eigenvalue $\lambda_i$  at the SIM $K$, which is slightly different from the definition in Ref.~\cite{song2017} (see more details in Section C of the SM [SM C]).
 The $z_{2}$ indicator is computed to be 1 and 0 for \ti~ and \wsm~[See Table~\ref{tab:weyls4}  in the SM C], respectively.

In the 3D insulating phase, the strong TI (STI) index $\nu_0$~\cite{PhysRevLett.98.106803} is defined on eight distinct TRI momenta [$\Gamma_{i=(n_1n_2n_3)}=(n_1\bb_1+n_2\bb_2+n_3\bb_3)/2$ with $n_j=0,1$ and $\bb_j$ primitive reciprocal lattice vectors]:
\begin{equation*}
(-1)^{\nu_0}=\prod_{n_j=0,1}  \delta_{n_1n_2n_3}=(-1)^{\nu_{a_1}}(-1)^{\nu_{a_2}},
\end{equation*}
where $\delta_i=\sqrt{det[w(\Gamma_i)]}/Pf[w(\Gamma_i)]$ with the unitary matrix $w_{ij}(\bk)=\braket{u_i(\bk)}{{\cal T}|u_j(\bk)}$. Here $\ket{u_j(\bk)}$ is the periodic part of the Bloch wavefunction. At $\bk=\Gamma_i$, $w_{ij} =-w_{ji}$, so the Pfaffian $Pf[w(\Gamma_i)]$ is well defined.
Since the product of four $\delta_i$ defines the $\mathbb Z_2$ invariant of the 2D TRI plane (if the four $\Gamma_i$ form a plane), the STI indicator $\nu_0$ can be also defined by two distinct $\mathbb Z_2$ invariants ($\nu_{a_1}$ and $\nu_{a_2}$) in the two parallel TRI planes ($a_1$-plane and $a_2$-plane), which yields $\eta=\nu_0$ for an insulator. Note that $\nu_0$ is well defined if the 3D bulk states are fully gapped while $\eta$ is well defined as long as there are two fully gapped TRI planes. On the other hand, in the presence of an additional symmetry $S_4$, the $z_2$ indicator of $S_4$ symmetry is presented to be identical to the STI $\mathbb Z_2$ indicator $\nu_0$ for the insulators~\cite{song2017,Haruki2018}.
Therefore, the ``TIs" in \wsm~in space group 121 can \emph{not} be insulators. Consequently, we find that those candidates are virtually WSMs, with four pairs of Weyl points being at the charge neutrality level.

\subsection{Weyl points and Wilson loop calculations}
To locate the positions of the Weyl points, we have calculated the $k_z$-directed Wilson loops of the WSM \wsmc~along the path: $\bar M [0.5,0.5]-\bar \Gamma[0,0]-\bar M'[0.5,-0.5]-\bar M[0.5,0.5]$ (in units of $[\frac{2\pi}{a},\frac{2\pi}{a}]$) in the $k_xk_y$ plane. The results in Fig.~\ref{fig:3}(a) show that it has a nontrivial Chern number $C=+2$, which implies that at least two Weyl points with charge +1 are enclosed in the 2D manifold spanned by the in-plane path and $k_z$ axis~\cite{Fang92,Balents2011Weyl}.
First, let's assume there is a Weyl point with charge +1 at a general point [$x_1 (\frac{2\pi}{a}),y_1 (\frac{2\pi}{a}),z_1 (\frac{2\pi}{c})$] in the manifold. Since it's fully gapped in the $k_z=0$ plane, $z_1$ should be nonzero (\ie $z_1\neq 0$).
Then, the combined symmetry ${\cal T}C_{2z}$ yields that there is also a Weyl point at [$x_1,y_1,-z_1$] with the same chiral charge $+1$.
Lastly, if the Weyl points are away from the $k_y=0$ plane, the number of the Weyl points enclosed in the manifold must be a multiple of four with the same topological chiral charge due to the two antiunitary symmetries: ${\cal T}C_{2y}$ and ${\cal T}C_{2z}$. Thus, the corresponding Chern number along the path has to be a multiple of four.
But, it's obvious not the case in this compound. Therefore, we conjecture the Weyl points are located in the $k_y=0$ plane: ($x_1,0,\pm z_1$). After carefully checking the energy gap and topological chiral charge in half of the $k_y=0$ plane (\ie $k_x>0$), we do find two Weyl points at [$0.0036,0.0,\pm0.0657$]. The topological chiral charge is computed with the Wilson-loop method on an enclosed manifold surrounding the Weyl point. The results of the Weyl point [$0.0036,0.0,0.0657$] are shown in Fig.~\ref{fig:3}(b) and its topological charge is read to be $+1$. Considering the two Weyl points with the same chiral charge, it is consistent with the total Chern number ($C=+2$) in Fig.~\ref{fig:3}(a). 

\begin{table*}[!htb]
\begin{tabular}{c|c|c|c|c|c|c|c|c|c|c|c|c}
\hline
\hline
 Phases  & $A_0$ & A$_1$ & A$_2$ & $B_0$ & B$_1$ & B$_2$ & C$_1$ & C$_2$ & C$_3$ &$\delta_1$ &$\delta_2$ &$\delta_3$ \\
    & (eV) & (eV$\cdot \AA^2$) & (eV$\cdot \AA^2$) &  (eV) & (eV$\cdot \AA^2$) & (eV$\cdot \AA^2$) &  (eV$\cdot \AA^2$) & (eV$\cdot \AA^2$) & (eV$\cdot \AA$) & (eV$\cdot \AA$) &(eV$\cdot \AA$)&(eV$\cdot \AA^3$) \\
\hline
 TI      & -0.055 & 25.121 & 28.679 & -0.001 & -6.642 & -2.872 & 0.244 & 4.691 &  0.325 & 0.020 & 0.013 & 1.103\\
WSM      & -0.151 & 27.895 & 18.702 & -0.020 & -5.451 & -2.369 & 0.300 & 3.300 &  1.137 & -0.034 & 1.300 & 4.400\\
\hline
\hline
\end{tabular}
\caption{The six-band model fitting parameters for the \tic~(Case I) and \wsmc~(Case II) compounds.
}\label{tab:matpara}
\end{table*}

\subsection{Effective model and Fermi arcs}
To capture the nontrivial topology of these compounds, we build a six-band effective model, which includes four valence bands ($\Gamma_6$ and $\Gamma_7$) and two conduction bands ($\Gamma_6$). Under the basis of $\{i|xyz\up\rangle,i|xyz\dw\rangle, |\frac{3}{2}, \frac{3}{2}\rangle, |\frac{3}{2}, \frac{1}{2}\rangle, |\frac{3}{2}, -\frac{1}{2}\rangle, |\frac{3}{2}, -\frac{3}{2}\rangle\}$, the $D_{2d}$-invariant $\bold{k\cdot p}$ Hamiltonian can be given as follows:
\begin{equation*}
    \begin{split}
        &H(\bk) = \begin{bmatrix}
            M_0& C_3{\mathbb S}^\dagger \\
            C_3{\mathbb S} & H_0+\delta_1 H_A+\delta_2 H_B +\delta_3H_C
        \end{bmatrix}
    \end{split}
\end{equation*}
where $M_0=\left(A_0+A_1k_z^2+A_2k_{||}^2\right) {\mathbb I}_{2}$ and  $H_0=\left(B_0+B_1k_z^2+B_2k_{||}^2\right){\mathbb I}_{4}+C_1 {\mathbb E}+C_2{\mathbb T}$ ( $\mathbb I_n$ is the $n\times n$ identity matrix, and see the explicit matrices of $\mathbb E$, $\mathbb T$, $\mathbb S$ and $H_C$ in Section D of the SM), $H_A= diag\{1,-1,-1,1\}$, and
\begin{equation*}
  H_B = \begin{pmatrix}
      0 & -k_+ & 2k_z & -\sqrt{3}k_- \\
      -k_- & 0 & \sqrt{3}k_+ & -2k_z \\
      2k_z & \sqrt{3}k_- & 0 & -k_+ \\
      -\sqrt{3}k_+ & -2k_z & -k_- & 0
  \end{pmatrix}
\end{equation*}
 When $A_1=A_2$, $B_1=B_2$, $\delta_1=\delta_2=\delta_3=0$, it's actually $O_h$-invariant. The $H_A$ term is a uni-axial strain, which reduces the symmetry to $D_{4h}$. The $H_B$ term is critical, which breaks both $I$ and $C_{4z}$, but keeps $S_{4z}$.
The $A_{1,2}>0$ and $B_{1,2}<0$ stand for the four valence bands and two conduction bands in the origin ($A_0>B_0$).
The $A_0<B_0$ represents the band inversion happening at the $\Gamma$ point. As a result, the $k_z=0$ plane has a nontrivial $\mathbb Z_2$ invariant with four occupied bands (\ie $\nu_{k_z=0}=1$). 
If $\delta_1>0$, it's a TI without gapless points. If $\delta_1<0$, it's a WSM with four pairs of Weyl points.
The fitting parameters for~\tic~and~\wsmc~are given in Table~\ref{tab:matpara} and the corresponding band structures can be found in Fig.~\ref{fig:s3}.

\begin{figure}[!htb]
\centering
\includegraphics[width=8.5 cm]{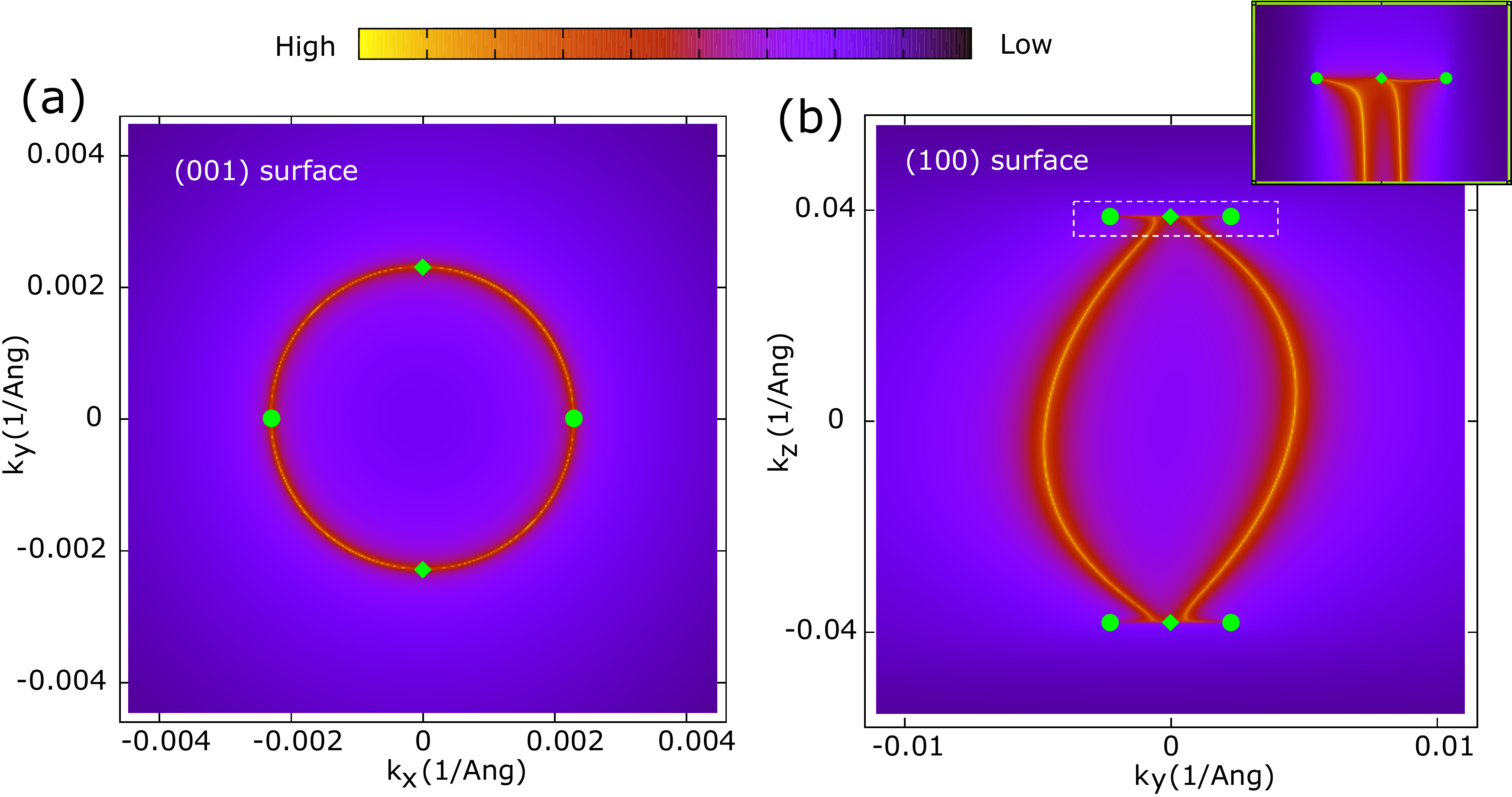}
\caption{(Color online)
Surface Fermi arcs of six-band model. (a) Surface Fermi arcs in the (001) surface BZ. (b) Surface Fermi arcs in the (100) surface BZ. The projected Weyl points are shown as square and circle points for different chiralities.
}
\label{fig:4}
\end{figure}

To obtain the Fermi-arc~\cite{Balents2011Weyl,Xu2011Chern} states of the WSM in \wsm, we transform the six-band model into a tight-binding model on a tetragonal lattice
by introducing the substitutions: $k_{i}\rightarrow \frac{1}{L_{i}}\text{sin}[k_{i}L_{i}]$ and $k_{i}^2\rightarrow\frac{2}{L_{i}^2}(1-\text{cos}[k_{i}L_{i}])$ with $i=x,y,z$~\cite{Wang2013}.
We use an iterative method to obtain the surface Green's function of the semi-infinite system~\cite{Wu2017WannierTools,Sancho_1985}. The imaginary part of the surface Green's function is the local density of states (LDOS) at the surface. The obtained LDOS on semi-infinite (001) and (100) surfaces are presented in Fig.~\ref{fig:4}. Since the Weyl points are exactly located at the charge neutrality level, we only see the Fermi-arc states connecting the projections of the Weyl points. For the (001) surface, two same-chirality Weyl points project onto the same projection, so each projection has two arc states. For the (100) surface, the projected topological charges are presented in Fig.~\ref{fig:4}(b), two arc states have to go across the $k_z=0$ line, because it's the edge of the $k_z=0$ plane with a nontrivial
$\mathbb Z_2$ invariant.

\section{Discussion}
To check the stability of the band inversion in these WSMs (see Table \ref{tab:s1} in the SM A) in space group 121, we have performed more accurate calculations by using a modified Becke-Johnson (mBJ) potential. The evolutions of the energy levels of the four bands at $\Gamma$ (three valence bands and one conduction band) are presented in Section B of the SM as a function of the mBJ parameter ($C_\text{mBJ}$). The results show that the band inversion feature for Cu$_2$HgGeTe$_4$, Cu$_3$SbSe$_4$, Cu$_2$HgSnSe$_4$, and   Cu$_2$HgSnTe$_4$ are relatively more reliable. Among them, Cu$_2$HgSnTe$_4$ is the most promising candidate in the WSM phase, with the band inversion happened at a relatively large $C_\text{mBJ}=1.25$.

The criterion of $\eta\neq z_2$ for a WSM phase can be widely applied to other space groups with $S_4$ symmetry. For example, we have computed the indicator $z_2$ and the invariant $\eta$ of the WSM CuTlSe$_2$ of space group 122~\cite{Haijun2016}, which was also previously predicted to be a TI~\cite{Feng2011}. The obtained results of $\eta=1$ and $z_2=0$ are consistent with the WSM phase. In addition, this criterion can be used to understand the robustness of the WSM phase (compressive strain) and the TI phase (tensile strain) of the strained HgTe material as well~\cite{HgTenc2016}.

In summary, based on the DFT calculations,
we demonstrate that the previously predicted ``TIs" in space group 121,
which are lack of inversion symmetry but respect $S_{4z}$ symmetry, can be actually classified into two cases: $z_2=1 $ (\ti) and $z_2=0$ (\wsm).
The common characteristic of these ``TIs" is that the time-reversal $\mathbb Z_2$ invariants are 1 and 0 for the $k_z=0$ and $k_z=\frac{\pi}{c}$ planes, respectively, resulting in $\eta=1$.
It's consistent with $S_4~z_2=1$ in \ti~for an insulating phase.
But, the ``TIs" with $S_4~z_2=0$ in \wsm~are actually WSMs, which are not revealed before.
They are also serving as typical examples of topological materials with \emph{trivial} symmetry indicators~\cite{song2017,wangprl2019}.
Four pairs of Weyl points are found in the $k_{x,y}=0$ planes, with each plane having four Weyl points with the same topological chiral charge.
Our work corrects the topological knowledge of these compounds and predicts more WSM candidates which can be further checked in experiments.
More importantly, the strategy to find the Weyl points in the TRI systems with symmetry indicators and invariants (\ie $\eta\neq z_2$) opens a new route to search for WSMs, which could largely stimulate the prediction of the WSMs.

\ \\
\noindent \textbf{Acknowledgments}
We thank Prof. Hai-Jun Zhang for helpful discussions.
This work was supported by the National Natural Science Foundation of China  (11504117, 11674369, 11974395 ). H.W. acknowledges support from the National Key Research and Development Program of China (Grant Nos. 2016YFA0300600, and 2018YFA0305700), the K. C. Wong Education Foundation (GJTD-2018-01). Z.W. acknowledges support from the National Thousand-Young-Talents Program and the CAS Pioneer Hundred Talents Program.

\bibliography{tps}

\begin{thebibliography}{60}
\expandafter\ifx\csname natexlab\endcsname\relax\def\natexlab#1{#1}\fi
\expandafter\ifx\csname bibnamefont\endcsname\relax
  \def\bibnamefont#1{#1}\fi
\expandafter\ifx\csname bibfnamefont\endcsname\relax
  \def\bibfnamefont#1{#1}\fi
\expandafter\ifx\csname citenamefont\endcsname\relax
  \def\citenamefont#1{#1}\fi
\expandafter\ifx\csname url\endcsname\relax
  \def\url#1{\texttt{#1}}\fi
\expandafter\ifx\csname urlprefix\endcsname\relax\def\urlprefix{URL }\fi
\providecommand{\bibinfo}[2]{#2}
\providecommand{\eprint}[2][]{\url{#2}}

\bibitem[{\citenamefont{Armitage et~al.}(2018)\citenamefont{Armitage, Mele, and
  Vishwanath}}]{RevModPhys.90.015001}
\bibinfo{author}{\bibfnamefont{N.~P.} \bibnamefont{Armitage}},
  \bibinfo{author}{\bibfnamefont{E.~J.} \bibnamefont{Mele}}, \bibnamefont{and}
  \bibinfo{author}{\bibfnamefont{A.}~\bibnamefont{Vishwanath}},
  \bibinfo{journal}{Rev. Mod. Phys.} \textbf{\bibinfo{volume}{90}},
  \bibinfo{pages}{015001} (\bibinfo{year}{2018}).

\bibitem[{\citenamefont{Wan et~al.}(2011)\citenamefont{Wan, Turner, Vishwanath,
  and Savrasov}}]{Wan2011}
\bibinfo{author}{\bibfnamefont{X.}~\bibnamefont{Wan}},
  \bibinfo{author}{\bibfnamefont{A.~M.} \bibnamefont{Turner}},
  \bibinfo{author}{\bibfnamefont{A.}~\bibnamefont{Vishwanath}},
  \bibnamefont{and} \bibinfo{author}{\bibfnamefont{S.~Y.}
  \bibnamefont{Savrasov}}, \bibinfo{journal}{Phys. Rev. B}
  \textbf{\bibinfo{volume}{83}}, \bibinfo{pages}{205101}
  (\bibinfo{year}{2011}).

\bibitem[{\citenamefont{Weng et~al.}(2016{\natexlab{a}})\citenamefont{Weng,
  Dai, and Fang}}]{Weng2016Topological}
\bibinfo{author}{\bibfnamefont{H.}~\bibnamefont{Weng}},
  \bibinfo{author}{\bibfnamefont{X.}~\bibnamefont{Dai}}, \bibnamefont{and}
  \bibinfo{author}{\bibfnamefont{Z.}~\bibnamefont{Fang}}, \bibinfo{journal}{J
  Phys Condens Matter} \textbf{\bibinfo{volume}{28}}, \bibinfo{pages}{303001}
  (\bibinfo{year}{2016}{\natexlab{a}}).

\bibitem[{\citenamefont{Wang et~al.}(2013)\citenamefont{Wang, Weng, Wu, Dai,
  and Fang}}]{Wang2013}
\bibinfo{author}{\bibfnamefont{Z.}~\bibnamefont{Wang}},
  \bibinfo{author}{\bibfnamefont{H.}~\bibnamefont{Weng}},
  \bibinfo{author}{\bibfnamefont{Q.}~\bibnamefont{Wu}},
  \bibinfo{author}{\bibfnamefont{X.}~\bibnamefont{Dai}}, \bibnamefont{and}
  \bibinfo{author}{\bibfnamefont{Z.}~\bibnamefont{Fang}},
  \bibinfo{journal}{Phys. Rev. B} \textbf{\bibinfo{volume}{88}},
  \bibinfo{pages}{125427} (\bibinfo{year}{2013}).

\bibitem[{\citenamefont{Qi and Zhang}(2010)}]{Qi2010The}
\bibinfo{author}{\bibfnamefont{X.~L.} \bibnamefont{Qi}} \bibnamefont{and}
  \bibinfo{author}{\bibfnamefont{S.-C.} \bibnamefont{Zhang}},
  \bibinfo{journal}{Physics Today} \textbf{\bibinfo{volume}{63}},
  \bibinfo{pages}{33} (\bibinfo{year}{2010}).

\bibitem[{\citenamefont{Hasan and Kane}(2010)}]{Hasan2010Topological}
\bibinfo{author}{\bibfnamefont{M.~Z.} \bibnamefont{Hasan}} \bibnamefont{and}
  \bibinfo{author}{\bibfnamefont{C.~L.} \bibnamefont{Kane}},
  \bibinfo{journal}{Rev. Mod. Phys.} \textbf{\bibinfo{volume}{82}},
  \bibinfo{pages}{3045} (\bibinfo{year}{2010}).

\bibitem[{\citenamefont{Bernevig et~al.}(2006)\citenamefont{Bernevig, Hughes,
  and Zhang}}]{Bernevig2006Quantum}
\bibinfo{author}{\bibfnamefont{B.~A.} \bibnamefont{Bernevig}},
  \bibinfo{author}{\bibfnamefont{T.~L.} \bibnamefont{Hughes}},
  \bibnamefont{and} \bibinfo{author}{\bibfnamefont{S.-C.} \bibnamefont{Zhang}},
  \bibinfo{journal}{Science} \textbf{\bibinfo{volume}{314}},
  \bibinfo{pages}{1757} (\bibinfo{year}{2006}).

\bibitem[{\citenamefont{Zhang et~al.}(2009)\citenamefont{Zhang, Liu, Qi, Dai,
  Fang, and Zhang}}]{Zhang2009Topological}
\bibinfo{author}{\bibfnamefont{H.}~\bibnamefont{Zhang}},
  \bibinfo{author}{\bibfnamefont{C.-X.} \bibnamefont{Liu}},
  \bibinfo{author}{\bibfnamefont{X.-L.} \bibnamefont{Qi}},
  \bibinfo{author}{\bibfnamefont{X.}~\bibnamefont{Dai}},
  \bibinfo{author}{\bibfnamefont{Z.}~\bibnamefont{Fang}}, \bibnamefont{and}
  \bibinfo{author}{\bibfnamefont{S.-C.} \bibnamefont{Zhang}},
  \bibinfo{journal}{Nature Physics} \textbf{\bibinfo{volume}{5}},
  \bibinfo{pages}{438} (\bibinfo{year}{2009}).

\bibitem[{\citenamefont{Cano et~al.}(2017)\citenamefont{Cano, Bradlyn, Wang,
  Hirschberger, Ong, and Bernevig}}]{PRBwzj}
\bibinfo{author}{\bibfnamefont{J.}~\bibnamefont{Cano}},
  \bibinfo{author}{\bibfnamefont{B.}~\bibnamefont{Bradlyn}},
  \bibinfo{author}{\bibfnamefont{Z.}~\bibnamefont{Wang}},
  \bibinfo{author}{\bibfnamefont{M.}~\bibnamefont{Hirschberger}},
  \bibinfo{author}{\bibfnamefont{N.~P.} \bibnamefont{Ong}}, \bibnamefont{and}
  \bibinfo{author}{\bibfnamefont{B.~A.} \bibnamefont{Bernevig}},
  \bibinfo{journal}{Phys. Rev. B} \textbf{\bibinfo{volume}{95}},
  \bibinfo{pages}{161306} (\bibinfo{year}{2017}).

\bibitem[{\citenamefont{Wang et~al.}(2016{\natexlab{a}})\citenamefont{Wang,
  Alexandradinata, Cava, and Bernevig}}]{Wang2016}
\bibinfo{author}{\bibfnamefont{Z.}~\bibnamefont{Wang}},
  \bibinfo{author}{\bibfnamefont{A.}~\bibnamefont{Alexandradinata}},
  \bibinfo{author}{\bibfnamefont{R.~J.} \bibnamefont{Cava}}, \bibnamefont{and}
  \bibinfo{author}{\bibfnamefont{B.~A.} \bibnamefont{Bernevig}},
  \bibinfo{journal}{Nature} \textbf{\bibinfo{volume}{532}},
  \bibinfo{pages}{189} (\bibinfo{year}{2016}{\natexlab{a}}).

\bibitem[{\citenamefont{K{\"o}nig et~al.}(2007)\citenamefont{K{\"o}nig,
  Wiedmann, Br{\"u}ne, Roth, Buhmann, Molenkamp, Qi, and
  Zhang}}]{konig2007quantum}
\bibinfo{author}{\bibfnamefont{M.}~\bibnamefont{K{\"o}nig}},
  \bibinfo{author}{\bibfnamefont{S.}~\bibnamefont{Wiedmann}},
  \bibinfo{author}{\bibfnamefont{C.}~\bibnamefont{Br{\"u}ne}},
  \bibinfo{author}{\bibfnamefont{A.}~\bibnamefont{Roth}},
  \bibinfo{author}{\bibfnamefont{H.}~\bibnamefont{Buhmann}},
  \bibinfo{author}{\bibfnamefont{L.~W.} \bibnamefont{Molenkamp}},
  \bibinfo{author}{\bibfnamefont{X.-L.} \bibnamefont{Qi}}, \bibnamefont{and}
  \bibinfo{author}{\bibfnamefont{S.-C.} \bibnamefont{Zhang}},
  \bibinfo{journal}{Science} \textbf{\bibinfo{volume}{318}},
  \bibinfo{pages}{766} (\bibinfo{year}{2007}).

\bibitem[{\citenamefont{Chen et~al.}(2009)\citenamefont{Chen, Analytis, Chu,
  Liu, Mo, Qi, Zhang, Lu, Dai, Fang et~al.}}]{chen2009experimental}
\bibinfo{author}{\bibfnamefont{Y.}~\bibnamefont{Chen}},
  \bibinfo{author}{\bibfnamefont{J.~G.} \bibnamefont{Analytis}},
  \bibinfo{author}{\bibfnamefont{J.-H.} \bibnamefont{Chu}},
  \bibinfo{author}{\bibfnamefont{Z.}~\bibnamefont{Liu}},
  \bibinfo{author}{\bibfnamefont{S.-K.} \bibnamefont{Mo}},
  \bibinfo{author}{\bibfnamefont{X.-L.} \bibnamefont{Qi}},
  \bibinfo{author}{\bibfnamefont{H.}~\bibnamefont{Zhang}},
  \bibinfo{author}{\bibfnamefont{D.}~\bibnamefont{Lu}},
  \bibinfo{author}{\bibfnamefont{X.}~\bibnamefont{Dai}},
  \bibinfo{author}{\bibfnamefont{Z.}~\bibnamefont{Fang}}, \bibnamefont{et~al.},
  \bibinfo{journal}{science} \textbf{\bibinfo{volume}{325}},
  \bibinfo{pages}{178} (\bibinfo{year}{2009}).

\bibitem[{\citenamefont{Ma et~al.}(2017)\citenamefont{Ma, Yi, Lv, Wang, Nie,
  Wang, Kong, Huang, Richard, Zhang et~al.}}]{advwang}
\bibinfo{author}{\bibfnamefont{J.}~\bibnamefont{Ma}},
  \bibinfo{author}{\bibfnamefont{C.}~\bibnamefont{Yi}},
  \bibinfo{author}{\bibfnamefont{B.}~\bibnamefont{Lv}},
  \bibinfo{author}{\bibfnamefont{Z.}~\bibnamefont{Wang}},
  \bibinfo{author}{\bibfnamefont{S.}~\bibnamefont{Nie}},
  \bibinfo{author}{\bibfnamefont{L.}~\bibnamefont{Wang}},
  \bibinfo{author}{\bibfnamefont{L.}~\bibnamefont{Kong}},
  \bibinfo{author}{\bibfnamefont{Y.}~\bibnamefont{Huang}},
  \bibinfo{author}{\bibfnamefont{P.}~\bibnamefont{Richard}},
  \bibinfo{author}{\bibfnamefont{P.}~\bibnamefont{Zhang}},
  \bibnamefont{et~al.}, \bibinfo{journal}{Science Advances}
  \textbf{\bibinfo{volume}{3}} (\bibinfo{year}{2017}).

\bibitem[{\citenamefont{Fu et~al.}(2007)\citenamefont{Fu, Kane, and
  Mele}}]{PhysRevLett.98.106803}
\bibinfo{author}{\bibfnamefont{L.}~\bibnamefont{Fu}},
  \bibinfo{author}{\bibfnamefont{C.~L.} \bibnamefont{Kane}}, \bibnamefont{and}
  \bibinfo{author}{\bibfnamefont{E.~J.} \bibnamefont{Mele}},
  \bibinfo{journal}{Phys. Rev. Lett.} \textbf{\bibinfo{volume}{98}},
  \bibinfo{pages}{106803} (\bibinfo{year}{2007}).

\bibitem[{\citenamefont{Fu and Kane}(2007)}]{Fu2007IS}
\bibinfo{author}{\bibfnamefont{L.}~\bibnamefont{Fu}} \bibnamefont{and}
  \bibinfo{author}{\bibfnamefont{C.~L.} \bibnamefont{Kane}},
  \bibinfo{journal}{Phys. Rev. B} \textbf{\bibinfo{volume}{76}},
  \bibinfo{pages}{045302} (\bibinfo{year}{2007}).

\bibitem[{\citenamefont{Khalaf et~al.}(2018)\citenamefont{Khalaf, Po,
  Vishwanath, and Watanabe}}]{Haruki2018}
\bibinfo{author}{\bibfnamefont{E.}~\bibnamefont{Khalaf}},
  \bibinfo{author}{\bibfnamefont{H.~C.} \bibnamefont{Po}},
  \bibinfo{author}{\bibfnamefont{A.}~\bibnamefont{Vishwanath}},
  \bibnamefont{and} \bibinfo{author}{\bibfnamefont{H.}~\bibnamefont{Watanabe}},
  \bibinfo{journal}{Phys. Rev. X} \textbf{\bibinfo{volume}{8}},
  \bibinfo{pages}{031070} (\bibinfo{year}{2018}).

\bibitem[{\citenamefont{{Song} et~al.}(2018)\citenamefont{{Song}, {Zhang},
  {Fang}, and {Fang}}}]{song2017}
\bibinfo{author}{\bibfnamefont{Z.}~\bibnamefont{{Song}}},
  \bibinfo{author}{\bibfnamefont{T.}~\bibnamefont{{Zhang}}},
  \bibinfo{author}{\bibfnamefont{Z.}~\bibnamefont{{Fang}}}, \bibnamefont{and}
  \bibinfo{author}{\bibfnamefont{C.}~\bibnamefont{{Fang}}},
  \bibinfo{journal}{Nature Communications} \textbf{\bibinfo{volume}{9}},
  \bibinfo{pages}{3530} (\bibinfo{year}{2018}).

\bibitem[{\citenamefont{Vergniory et~al.}(2019)\citenamefont{Vergniory, Elcoro,
  Felser, Regnault, Bernevig, and Wang}}]{nature_2019}
\bibinfo{author}{\bibfnamefont{M.~G.} \bibnamefont{Vergniory}},
  \bibinfo{author}{\bibfnamefont{L.}~\bibnamefont{Elcoro}},
  \bibinfo{author}{\bibfnamefont{C.}~\bibnamefont{Felser}},
  \bibinfo{author}{\bibfnamefont{N.}~\bibnamefont{Regnault}},
  \bibinfo{author}{\bibfnamefont{B.~A.} \bibnamefont{Bernevig}},
  \bibnamefont{and} \bibinfo{author}{\bibfnamefont{Z.}~\bibnamefont{Wang}},
  \bibinfo{journal}{Nature} \textbf{\bibinfo{volume}{566}},
  \bibinfo{pages}{480} (\bibinfo{year}{2019}).

\bibitem[{\citenamefont{Zhang et~al.}(2019)\citenamefont{Zhang, Jiang, Song,
  Huang, He, Fang, Weng, and Fang}}]{Zhang2018}
\bibinfo{author}{\bibfnamefont{T.}~\bibnamefont{Zhang}},
  \bibinfo{author}{\bibfnamefont{Y.}~\bibnamefont{Jiang}},
  \bibinfo{author}{\bibfnamefont{Z.}~\bibnamefont{Song}},
  \bibinfo{author}{\bibfnamefont{H.}~\bibnamefont{Huang}},
  \bibinfo{author}{\bibfnamefont{Y.}~\bibnamefont{He}},
  \bibinfo{author}{\bibfnamefont{Z.}~\bibnamefont{Fang}},
  \bibinfo{author}{\bibfnamefont{H.}~\bibnamefont{Weng}}, \bibnamefont{and}
  \bibinfo{author}{\bibfnamefont{C.}~\bibnamefont{Fang}},
  \bibinfo{journal}{Nature} \textbf{\bibinfo{volume}{566}},
  \bibinfo{pages}{475} (\bibinfo{year}{2019}).

\bibitem[{\citenamefont{Tang et~al.}(2019)\citenamefont{Tang, Po, Vishwanath,
  and Wan}}]{wanxg2019}
\bibinfo{author}{\bibfnamefont{F.}~\bibnamefont{Tang}},
  \bibinfo{author}{\bibfnamefont{H.~C.} \bibnamefont{Po}},
  \bibinfo{author}{\bibfnamefont{A.}~\bibnamefont{Vishwanath}},
  \bibnamefont{and} \bibinfo{author}{\bibfnamefont{X.}~\bibnamefont{Wan}},
  \bibinfo{journal}{Nature} \textbf{\bibinfo{volume}{566}},
  \bibinfo{pages}{486} (\bibinfo{year}{2019}).

\bibitem[{\citenamefont{Kruthoff et~al.}(2017)\citenamefont{Kruthoff, de~Boer,
  van Wezel, Kane, and Slager}}]{Slager2017}
\bibinfo{author}{\bibfnamefont{J.}~\bibnamefont{Kruthoff}},
  \bibinfo{author}{\bibfnamefont{J.}~\bibnamefont{de~Boer}},
  \bibinfo{author}{\bibfnamefont{J.}~\bibnamefont{van Wezel}},
  \bibinfo{author}{\bibfnamefont{C.~L.} \bibnamefont{Kane}}, \bibnamefont{and}
  \bibinfo{author}{\bibfnamefont{R.-J.} \bibnamefont{Slager}},
  \bibinfo{journal}{Phys. Rev. X} \textbf{\bibinfo{volume}{7}},
  \bibinfo{pages}{041069} (\bibinfo{year}{2017}).

\bibitem[{\citenamefont{Murakami}(2007)}]{Murakami_2007}
\bibinfo{author}{\bibfnamefont{S.}~\bibnamefont{Murakami}},
  \bibinfo{journal}{New Journal of Physics} \textbf{\bibinfo{volume}{9}},
  \bibinfo{pages}{356} (\bibinfo{year}{2007}).

\bibitem[{\citenamefont{Liu and Vanderbilt}(2014)}]{Liu2014Weyl}
\bibinfo{author}{\bibfnamefont{J.}~\bibnamefont{Liu}} \bibnamefont{and}
  \bibinfo{author}{\bibfnamefont{D.}~\bibnamefont{Vanderbilt}},
  \bibinfo{journal}{Physical Review B} \textbf{\bibinfo{volume}{90}},
  \bibinfo{pages}{155316} (\bibinfo{year}{2014}).

\bibitem[{\citenamefont{Weng et~al.}(2015)\citenamefont{Weng, Fang, Fang,
  Bernevig, and Dai}}]{wengprx}
\bibinfo{author}{\bibfnamefont{H.}~\bibnamefont{Weng}},
  \bibinfo{author}{\bibfnamefont{C.}~\bibnamefont{Fang}},
  \bibinfo{author}{\bibfnamefont{Z.}~\bibnamefont{Fang}},
  \bibinfo{author}{\bibfnamefont{B.~A.} \bibnamefont{Bernevig}},
  \bibnamefont{and} \bibinfo{author}{\bibfnamefont{X.}~\bibnamefont{Dai}},
  \bibinfo{journal}{Phys. Rev. X} \textbf{\bibinfo{volume}{5}},
  \bibinfo{pages}{011029} (\bibinfo{year}{2015}).

\bibitem[{\citenamefont{Soluyanov et~al.}(2015)\citenamefont{Soluyanov, Gresch,
  Wang, Wu, Troyer, Dai, and Bernevig}}]{Soluyanov2015Type}
\bibinfo{author}{\bibfnamefont{A.~A.} \bibnamefont{Soluyanov}},
  \bibinfo{author}{\bibfnamefont{D.}~\bibnamefont{Gresch}},
  \bibinfo{author}{\bibfnamefont{Z.}~\bibnamefont{Wang}},
  \bibinfo{author}{\bibfnamefont{Q.}~\bibnamefont{Wu}},
  \bibinfo{author}{\bibfnamefont{M.}~\bibnamefont{Troyer}},
  \bibinfo{author}{\bibfnamefont{X.}~\bibnamefont{Dai}}, \bibnamefont{and}
  \bibinfo{author}{\bibfnamefont{B.~A.} \bibnamefont{Bernevig}},
  \bibinfo{journal}{Nature} \textbf{\bibinfo{volume}{527}},
  \bibinfo{pages}{495} (\bibinfo{year}{2015}).

\bibitem[{\citenamefont{Weng et~al.}(2016{\natexlab{b}})\citenamefont{Weng,
  Fang, Fang, and Dai}}]{Weng2016Coexistence}
\bibinfo{author}{\bibfnamefont{H.}~\bibnamefont{Weng}},
  \bibinfo{author}{\bibfnamefont{C.}~\bibnamefont{Fang}},
  \bibinfo{author}{\bibfnamefont{Z.}~\bibnamefont{Fang}}, \bibnamefont{and}
  \bibinfo{author}{\bibfnamefont{X.}~\bibnamefont{Dai}},
  \bibinfo{journal}{Physical Review B} \textbf{\bibinfo{volume}{94}},
  \bibinfo{pages}{165201} (\bibinfo{year}{2016}{\natexlab{b}}).

\bibitem[{\citenamefont{Lv et~al.}(2015{\natexlab{a}})\citenamefont{Lv, Xu,
  Weng, Ma, Richard, Huang, Zhao, Chen, Matt, and Bisti}}]{Lv2015Observation}
\bibinfo{author}{\bibfnamefont{B.~Q.} \bibnamefont{Lv}},
  \bibinfo{author}{\bibfnamefont{N.}~\bibnamefont{Xu}},
  \bibinfo{author}{\bibfnamefont{H.~M.} \bibnamefont{Weng}},
  \bibinfo{author}{\bibfnamefont{J.~Z.} \bibnamefont{Ma}},
  \bibinfo{author}{\bibfnamefont{P.}~\bibnamefont{Richard}},
  \bibinfo{author}{\bibfnamefont{X.~C.} \bibnamefont{Huang}},
  \bibinfo{author}{\bibfnamefont{L.~X.} \bibnamefont{Zhao}},
  \bibinfo{author}{\bibfnamefont{G.~F.} \bibnamefont{Chen}},
  \bibinfo{author}{\bibfnamefont{C.~E.} \bibnamefont{Matt}}, \bibnamefont{and}
  \bibinfo{author}{\bibfnamefont{F.}~\bibnamefont{Bisti}},
  \bibinfo{journal}{Nature Physics} \textbf{\bibinfo{volume}{11}}
  (\bibinfo{year}{2015}{\natexlab{a}}).

\bibitem[{\citenamefont{Lv et~al.}(2015{\natexlab{b}})\citenamefont{Lv, Weng,
  Fu, Wang, Miao, Ma, Richard, Huang, Zhao, and Chen}}]{Lv2015Experimental}
\bibinfo{author}{\bibfnamefont{B.}~\bibnamefont{Lv}},
  \bibinfo{author}{\bibfnamefont{H.}~\bibnamefont{Weng}},
  \bibinfo{author}{\bibfnamefont{B.}~\bibnamefont{Fu}},
  \bibinfo{author}{\bibfnamefont{X.}~\bibnamefont{Wang}},
  \bibinfo{author}{\bibfnamefont{H.}~\bibnamefont{Miao}},
  \bibinfo{author}{\bibfnamefont{J.}~\bibnamefont{Ma}},
  \bibinfo{author}{\bibfnamefont{P.}~\bibnamefont{Richard}},
  \bibinfo{author}{\bibfnamefont{X.}~\bibnamefont{Huang}},
  \bibinfo{author}{\bibfnamefont{L.}~\bibnamefont{Zhao}}, \bibnamefont{and}
  \bibinfo{author}{\bibfnamefont{G.~a.} \bibnamefont{Chen}},
  \bibinfo{journal}{Physical Review X} \textbf{\bibinfo{volume}{5}},
  \bibinfo{pages}{031013} (\bibinfo{year}{2015}{\natexlab{b}}).

\bibitem[{\citenamefont{Nie et~al.}(2017)\citenamefont{Nie, Xu, Prinz, and
  Zhang}}]{nie2017topological}
\bibinfo{author}{\bibfnamefont{S.}~\bibnamefont{Nie}},
  \bibinfo{author}{\bibfnamefont{G.}~\bibnamefont{Xu}},
  \bibinfo{author}{\bibfnamefont{F.~B.} \bibnamefont{Prinz}}, \bibnamefont{and}
  \bibinfo{author}{\bibfnamefont{S.-C.} \bibnamefont{Zhang}},
  \bibinfo{journal}{Proceedings of the National Academy of Sciences}
  \textbf{\bibinfo{volume}{114}}, \bibinfo{pages}{10596}
  (\bibinfo{year}{2017}).

\bibitem[{\citenamefont{Wang et~al.}(2016{\natexlab{b}})\citenamefont{Wang,
  Vergniory, Kushwaha, Hirschberger, Chulkov, Ernst, Ong, Cava, and
  Bernevig}}]{PhysRevLett.117.236401}
\bibinfo{author}{\bibfnamefont{Z.}~\bibnamefont{Wang}},
  \bibinfo{author}{\bibfnamefont{M.~G.} \bibnamefont{Vergniory}},
  \bibinfo{author}{\bibfnamefont{S.}~\bibnamefont{Kushwaha}},
  \bibinfo{author}{\bibfnamefont{M.}~\bibnamefont{Hirschberger}},
  \bibinfo{author}{\bibfnamefont{E.~V.} \bibnamefont{Chulkov}},
  \bibinfo{author}{\bibfnamefont{A.}~\bibnamefont{Ernst}},
  \bibinfo{author}{\bibfnamefont{N.~P.} \bibnamefont{Ong}},
  \bibinfo{author}{\bibfnamefont{R.~J.} \bibnamefont{Cava}}, \bibnamefont{and}
  \bibinfo{author}{\bibfnamefont{B.~A.} \bibnamefont{Bernevig}},
  \bibinfo{journal}{Phys. Rev. Lett.} \textbf{\bibinfo{volume}{117}},
  \bibinfo{pages}{236401} (\bibinfo{year}{2016}{\natexlab{b}}).

\bibitem[{\citenamefont{Xu et~al.}(2016)\citenamefont{Xu, Weng, Lv, Matt, Park,
  Bisti, Strocov, Gawryluk, Pomjakushina, Conder et~al.}}]{xu2016observation}
\bibinfo{author}{\bibfnamefont{N.}~\bibnamefont{Xu}},
  \bibinfo{author}{\bibfnamefont{H.}~\bibnamefont{Weng}},
  \bibinfo{author}{\bibfnamefont{B.}~\bibnamefont{Lv}},
  \bibinfo{author}{\bibfnamefont{C.~E.} \bibnamefont{Matt}},
  \bibinfo{author}{\bibfnamefont{J.}~\bibnamefont{Park}},
  \bibinfo{author}{\bibfnamefont{F.}~\bibnamefont{Bisti}},
  \bibinfo{author}{\bibfnamefont{V.~N.} \bibnamefont{Strocov}},
  \bibinfo{author}{\bibfnamefont{D.}~\bibnamefont{Gawryluk}},
  \bibinfo{author}{\bibfnamefont{E.}~\bibnamefont{Pomjakushina}},
  \bibinfo{author}{\bibfnamefont{K.}~\bibnamefont{Conder}},
  \bibnamefont{et~al.}, \bibinfo{journal}{Nature communications}
  \textbf{\bibinfo{volume}{7}}, \bibinfo{pages}{11006} (\bibinfo{year}{2016}).

\bibitem[{\citenamefont{Xu et~al.}(2015)\citenamefont{Xu, Belopolski, Alidoust,
  Neupane, Bian, Zhang, Sankar, Chang, Yuan, Lee et~al.}}]{xu2015discovery}
\bibinfo{author}{\bibfnamefont{S.-Y.} \bibnamefont{Xu}},
  \bibinfo{author}{\bibfnamefont{I.}~\bibnamefont{Belopolski}},
  \bibinfo{author}{\bibfnamefont{N.}~\bibnamefont{Alidoust}},
  \bibinfo{author}{\bibfnamefont{M.}~\bibnamefont{Neupane}},
  \bibinfo{author}{\bibfnamefont{G.}~\bibnamefont{Bian}},
  \bibinfo{author}{\bibfnamefont{C.}~\bibnamefont{Zhang}},
  \bibinfo{author}{\bibfnamefont{R.}~\bibnamefont{Sankar}},
  \bibinfo{author}{\bibfnamefont{G.}~\bibnamefont{Chang}},
  \bibinfo{author}{\bibfnamefont{Z.}~\bibnamefont{Yuan}},
  \bibinfo{author}{\bibfnamefont{C.-C.} \bibnamefont{Lee}},
  \bibnamefont{et~al.}, \bibinfo{journal}{Science}
  \textbf{\bibinfo{volume}{349}}, \bibinfo{pages}{613} (\bibinfo{year}{2015}).

\bibitem[{\citenamefont{Wang et~al.}(2016{\natexlab{c}})\citenamefont{Wang,
  Zhang, Huang, Nie, Liu, Liang, Zhang, Shen, Liu, Hu
  et~al.}}]{wang2016observation2}
\bibinfo{author}{\bibfnamefont{C.}~\bibnamefont{Wang}},
  \bibinfo{author}{\bibfnamefont{Y.}~\bibnamefont{Zhang}},
  \bibinfo{author}{\bibfnamefont{J.}~\bibnamefont{Huang}},
  \bibinfo{author}{\bibfnamefont{S.}~\bibnamefont{Nie}},
  \bibinfo{author}{\bibfnamefont{G.}~\bibnamefont{Liu}},
  \bibinfo{author}{\bibfnamefont{A.}~\bibnamefont{Liang}},
  \bibinfo{author}{\bibfnamefont{Y.}~\bibnamefont{Zhang}},
  \bibinfo{author}{\bibfnamefont{B.}~\bibnamefont{Shen}},
  \bibinfo{author}{\bibfnamefont{J.}~\bibnamefont{Liu}},
  \bibinfo{author}{\bibfnamefont{C.}~\bibnamefont{Hu}}, \bibnamefont{et~al.},
  \bibinfo{journal}{Physical Review B} \textbf{\bibinfo{volume}{94}},
  \bibinfo{pages}{241119} (\bibinfo{year}{2016}{\natexlab{c}}).

\bibitem[{\citenamefont{Huang et~al.}(2015)\citenamefont{Huang, Zhao, Long,
  Wang, Chen, Yang, Liang, Xue, Weng, Fang et~al.}}]{huang2015observation}
\bibinfo{author}{\bibfnamefont{X.}~\bibnamefont{Huang}},
  \bibinfo{author}{\bibfnamefont{L.}~\bibnamefont{Zhao}},
  \bibinfo{author}{\bibfnamefont{Y.}~\bibnamefont{Long}},
  \bibinfo{author}{\bibfnamefont{P.}~\bibnamefont{Wang}},
  \bibinfo{author}{\bibfnamefont{D.}~\bibnamefont{Chen}},
  \bibinfo{author}{\bibfnamefont{Z.}~\bibnamefont{Yang}},
  \bibinfo{author}{\bibfnamefont{H.}~\bibnamefont{Liang}},
  \bibinfo{author}{\bibfnamefont{M.}~\bibnamefont{Xue}},
  \bibinfo{author}{\bibfnamefont{H.}~\bibnamefont{Weng}},
  \bibinfo{author}{\bibfnamefont{Z.}~\bibnamefont{Fang}}, \bibnamefont{et~al.},
  \bibinfo{journal}{Physical Review X} \textbf{\bibinfo{volume}{5}},
  \bibinfo{pages}{031023} (\bibinfo{year}{2015}).

\bibitem[{\citenamefont{Zhang et~al.}(2016)\citenamefont{Zhang, Xu, Belopolski,
  Yuan, Lin, Tong, Bian, Alidoust, Lee, Huang et~al.}}]{zhang2016signatures}
\bibinfo{author}{\bibfnamefont{C.-L.} \bibnamefont{Zhang}},
  \bibinfo{author}{\bibfnamefont{S.-Y.} \bibnamefont{Xu}},
  \bibinfo{author}{\bibfnamefont{I.}~\bibnamefont{Belopolski}},
  \bibinfo{author}{\bibfnamefont{Z.}~\bibnamefont{Yuan}},
  \bibinfo{author}{\bibfnamefont{Z.}~\bibnamefont{Lin}},
  \bibinfo{author}{\bibfnamefont{B.}~\bibnamefont{Tong}},
  \bibinfo{author}{\bibfnamefont{G.}~\bibnamefont{Bian}},
  \bibinfo{author}{\bibfnamefont{N.}~\bibnamefont{Alidoust}},
  \bibinfo{author}{\bibfnamefont{C.-C.} \bibnamefont{Lee}},
  \bibinfo{author}{\bibfnamefont{S.-M.} \bibnamefont{Huang}},
  \bibnamefont{et~al.}, \bibinfo{journal}{Nature communications}
  \textbf{\bibinfo{volume}{7}}, \bibinfo{pages}{10735} (\bibinfo{year}{2016}).

\bibitem[{\citenamefont{Xu et~al.}(2011)\citenamefont{Xu, Weng, Wang, Dai, and
  Fang}}]{Xu2011Chern}
\bibinfo{author}{\bibfnamefont{G.}~\bibnamefont{Xu}},
  \bibinfo{author}{\bibfnamefont{H.}~\bibnamefont{Weng}},
  \bibinfo{author}{\bibfnamefont{Z.}~\bibnamefont{Wang}},
  \bibinfo{author}{\bibfnamefont{X.}~\bibnamefont{Dai}}, \bibnamefont{and}
  \bibinfo{author}{\bibfnamefont{Z.}~\bibnamefont{Fang}},
  \bibinfo{journal}{Phys.Rev.lett} \textbf{\bibinfo{volume}{107}},
  \bibinfo{pages}{186806} (\bibinfo{year}{2011}).

\bibitem[{\citenamefont{Fang et~al.}(2003)\citenamefont{Fang, Nagaosa,
  Takahashi, Asamitsu, Mathieu, Ogasawara, Yamada, Kawasaki, Tokura, and
  Terakura}}]{Fang92}
\bibinfo{author}{\bibfnamefont{Z.}~\bibnamefont{Fang}},
  \bibinfo{author}{\bibfnamefont{N.}~\bibnamefont{Nagaosa}},
  \bibinfo{author}{\bibfnamefont{K.~S.} \bibnamefont{Takahashi}},
  \bibinfo{author}{\bibfnamefont{A.}~\bibnamefont{Asamitsu}},
  \bibinfo{author}{\bibfnamefont{R.}~\bibnamefont{Mathieu}},
  \bibinfo{author}{\bibfnamefont{T.}~\bibnamefont{Ogasawara}},
  \bibinfo{author}{\bibfnamefont{H.}~\bibnamefont{Yamada}},
  \bibinfo{author}{\bibfnamefont{M.}~\bibnamefont{Kawasaki}},
  \bibinfo{author}{\bibfnamefont{Y.}~\bibnamefont{Tokura}}, \bibnamefont{and}
  \bibinfo{author}{\bibfnamefont{K.}~\bibnamefont{Terakura}},
  \bibinfo{journal}{Science} \textbf{\bibinfo{volume}{302}},
  \bibinfo{pages}{92} (\bibinfo{year}{2003}).

\bibitem[{\citenamefont{Hughes et~al.}(2010)\citenamefont{Hughes, Prodan, and
  Bernevig}}]{Hughes2010Inversion}
\bibinfo{author}{\bibfnamefont{T.~L.} \bibnamefont{Hughes}},
  \bibinfo{author}{\bibfnamefont{E.}~\bibnamefont{Prodan}}, \bibnamefont{and}
  \bibinfo{author}{\bibfnamefont{B.~A.} \bibnamefont{Bernevig}},
  \bibinfo{journal}{Physical Review B Condensed Matter}
  \textbf{\bibinfo{volume}{83}}, \bibinfo{pages}{1417} (\bibinfo{year}{2010}).

\bibitem[{\citenamefont{Nie et~al.}(2019)\citenamefont{Nie, Sun, Prinz, Wang,
  Weng, Fang, and Dai}}]{nie2019magnetic}
\bibinfo{author}{\bibfnamefont{S.}~\bibnamefont{Nie}},
  \bibinfo{author}{\bibfnamefont{Y.}~\bibnamefont{Sun}},
  \bibinfo{author}{\bibfnamefont{F.~B.} \bibnamefont{Prinz}},
  \bibinfo{author}{\bibfnamefont{Z.}~\bibnamefont{Wang}},
  \bibinfo{author}{\bibfnamefont{H.}~\bibnamefont{Weng}},
  \bibinfo{author}{\bibfnamefont{Z.}~\bibnamefont{Fang}}, \bibnamefont{and}
  \bibinfo{author}{\bibfnamefont{X.}~\bibnamefont{Dai}},
  \bibinfo{journal}{arXiv preprint arXiv:1907.10051}  (\bibinfo{year}{2019}).

\bibitem[{\citenamefont{Gao et~al.}(work in progress)}]{tobedone2019}
\bibinfo{author}{\bibfnamefont{J.}~\bibnamefont{Gao}} \bibnamefont{et~al.}
  (\bibinfo{year}{work in progress}).

\bibitem[{\citenamefont{Kane and Mele}(2005)}]{Kane2005}
\bibinfo{author}{\bibfnamefont{C.~L.} \bibnamefont{Kane}} \bibnamefont{and}
  \bibinfo{author}{\bibfnamefont{E.~J.} \bibnamefont{Mele}},
  \bibinfo{journal}{Phys. Rev. Lett.} \textbf{\bibinfo{volume}{95}},
  \bibinfo{pages}{146802} (\bibinfo{year}{2005}).

\bibitem[{\citenamefont{Wang et~al.}(2010)\citenamefont{Wang, Lin, Das, Hasan,
  and Bansil}}]{Wang2010}
\bibinfo{author}{\bibfnamefont{Y.~J.} \bibnamefont{Wang}},
  \bibinfo{author}{\bibfnamefont{H.}~\bibnamefont{Lin}},
  \bibinfo{author}{\bibfnamefont{T.}~\bibnamefont{Das}},
  \bibinfo{author}{\bibfnamefont{M.~Z.} \bibnamefont{Hasan}}, \bibnamefont{and}
  \bibinfo{author}{\bibfnamefont{A.}~\bibnamefont{Bansil}},
  \bibinfo{journal}{New Journal of Physics} \textbf{\bibinfo{volume}{13}}
  (\bibinfo{year}{2010}).

\bibitem[{\citenamefont{Ruan et~al.}(2016{\natexlab{a}})\citenamefont{Ruan,
  Jian, Zhang, Yao, Zhang, Zhang, and Xing}}]{Haijun2016}
\bibinfo{author}{\bibfnamefont{J.}~\bibnamefont{Ruan}},
  \bibinfo{author}{\bibfnamefont{S.-K.} \bibnamefont{Jian}},
  \bibinfo{author}{\bibfnamefont{D.}~\bibnamefont{Zhang}},
  \bibinfo{author}{\bibfnamefont{H.}~\bibnamefont{Yao}},
  \bibinfo{author}{\bibfnamefont{H.}~\bibnamefont{Zhang}},
  \bibinfo{author}{\bibfnamefont{S.-C.} \bibnamefont{Zhang}}, \bibnamefont{and}
  \bibinfo{author}{\bibfnamefont{D.}~\bibnamefont{Xing}},
  \bibinfo{journal}{Phys. Rev. Lett.} \textbf{\bibinfo{volume}{116}},
  \bibinfo{pages}{226801} (\bibinfo{year}{2016}{\natexlab{a}}).

\bibitem[{\citenamefont{Ruan et~al.}(2016{\natexlab{b}})\citenamefont{Ruan,
  Jian, Yan, Zhang, Zhang, and Xing}}]{HgTenc2016}
\bibinfo{author}{\bibfnamefont{J.}~\bibnamefont{Ruan}},
  \bibinfo{author}{\bibfnamefont{S.-K.} \bibnamefont{Jian}},
  \bibinfo{author}{\bibfnamefont{H.}~\bibnamefont{Yan}},
  \bibinfo{author}{\bibfnamefont{H.}~\bibnamefont{Zhang}},
  \bibinfo{author}{\bibfnamefont{S.-C.} \bibnamefont{Zhang}}, \bibnamefont{and}
  \bibinfo{author}{\bibfnamefont{D.}~\bibnamefont{Xing}},
  \bibinfo{journal}{Nat. Commun.} \textbf{\bibinfo{volume}{7}},
  \bibinfo{pages}{11136} (\bibinfo{year}{2016}{\natexlab{b}}).

\bibitem[{\citenamefont{Kresse and Furthmüller}(1996)}]{KRESSE199615}
\bibinfo{author}{\bibfnamefont{G.}~\bibnamefont{Kresse}} \bibnamefont{and}
  \bibinfo{author}{\bibfnamefont{J.}~\bibnamefont{Furthmüller}},
  \bibinfo{journal}{Computational Materials Science}
  \textbf{\bibinfo{volume}{6}}, \bibinfo{pages}{15 } (\bibinfo{year}{1996}).

\bibitem[{\citenamefont{Kresse and Furthm\"uller}(1996)}]{vasp}
\bibinfo{author}{\bibfnamefont{G.}~\bibnamefont{Kresse}} \bibnamefont{and}
  \bibinfo{author}{\bibfnamefont{J.}~\bibnamefont{Furthm\"uller}},
  \bibinfo{journal}{Phys. Rev. B} \textbf{\bibinfo{volume}{54}},
  \bibinfo{pages}{11169} (\bibinfo{year}{1996}).

\bibitem[{\citenamefont{Bl\"ochl}(1994)}]{paw1}
\bibinfo{author}{\bibfnamefont{P.~E.} \bibnamefont{Bl\"ochl}},
  \bibinfo{journal}{Phys. Rev. B} \textbf{\bibinfo{volume}{50}},
  \bibinfo{pages}{17953} (\bibinfo{year}{1994}).

\bibitem[{\citenamefont{Kresse and Joubert}(1999)}]{paw2}
\bibinfo{author}{\bibfnamefont{G.}~\bibnamefont{Kresse}} \bibnamefont{and}
  \bibinfo{author}{\bibfnamefont{D.}~\bibnamefont{Joubert}},
  \bibinfo{journal}{Phys. Rev. B} \textbf{\bibinfo{volume}{59}},
  \bibinfo{pages}{1758} (\bibinfo{year}{1999}).

\bibitem[{\citenamefont{Perdew et~al.}(1996)\citenamefont{Perdew, Burke, and
  Ernzerhof}}]{pbe}
\bibinfo{author}{\bibfnamefont{J.~P.} \bibnamefont{Perdew}},
  \bibinfo{author}{\bibfnamefont{K.}~\bibnamefont{Burke}}, \bibnamefont{and}
  \bibinfo{author}{\bibfnamefont{M.}~\bibnamefont{Ernzerhof}},
  \bibinfo{journal}{Phys. Rev. Lett.} \textbf{\bibinfo{volume}{77}},
  \bibinfo{pages}{3865} (\bibinfo{year}{1996}).

\bibitem[{\citenamefont{Yu et~al.}(2011)\citenamefont{Yu, Qi, Bernevig, Fang,
  and Dai}}]{Yu2011An}
\bibinfo{author}{\bibfnamefont{R.}~\bibnamefont{Yu}},
  \bibinfo{author}{\bibfnamefont{X.~L.} \bibnamefont{Qi}},
  \bibinfo{author}{\bibfnamefont{A.}~\bibnamefont{Bernevig}},
  \bibinfo{author}{\bibfnamefont{Z.}~\bibnamefont{Fang}}, \bibnamefont{and}
  \bibinfo{author}{\bibfnamefont{X.}~\bibnamefont{Dai}},
  \bibinfo{journal}{Physical Review B Condensed Matter}
  \textbf{\bibinfo{volume}{84}}, \bibinfo{pages}{2250} (\bibinfo{year}{2011}).

\bibitem[{\citenamefont{Balents}(2011)}]{Balents2011Weyl}
\bibinfo{author}{\bibfnamefont{L.}~\bibnamefont{Balents}},
  \bibinfo{journal}{Physics} \textbf{\bibinfo{volume}{4}}
  (\bibinfo{year}{2011}).

\bibitem[{\citenamefont{Wu et~al.}(2018)\citenamefont{Wu, Zhang, Song, Troyer,
  and Soluyanov}}]{Wu2017WannierTools}
\bibinfo{author}{\bibfnamefont{Q.}~\bibnamefont{Wu}},
  \bibinfo{author}{\bibfnamefont{S.}~\bibnamefont{Zhang}},
  \bibinfo{author}{\bibfnamefont{H.-F.} \bibnamefont{Song}},
  \bibinfo{author}{\bibfnamefont{M.}~\bibnamefont{Troyer}}, \bibnamefont{and}
  \bibinfo{author}{\bibfnamefont{A.~A.} \bibnamefont{Soluyanov}},
  \bibinfo{journal}{Computer Physics Communications}
  \textbf{\bibinfo{volume}{224}}, \bibinfo{pages}{405} (\bibinfo{year}{2018}).

\bibitem[{\citenamefont{Sancho et~al.}(1985)\citenamefont{Sancho, Sancho,
  Sancho, and Rubio}}]{Sancho_1985}
\bibinfo{author}{\bibfnamefont{M.~P.~L.} \bibnamefont{Sancho}},
  \bibinfo{author}{\bibfnamefont{J.~M.~L.} \bibnamefont{Sancho}},
  \bibinfo{author}{\bibfnamefont{J.~M.~L.} \bibnamefont{Sancho}},
  \bibnamefont{and} \bibinfo{author}{\bibfnamefont{J.}~\bibnamefont{Rubio}},
  \bibinfo{journal}{Journal of Physics F: Metal Physics}
  \textbf{\bibinfo{volume}{15}}, \bibinfo{pages}{851} (\bibinfo{year}{1985}).

\bibitem[{\citenamefont{Feng et~al.}(2011)\citenamefont{Feng, Xiao, Ding, and
  Yao}}]{Feng2011}
\bibinfo{author}{\bibfnamefont{W.}~\bibnamefont{Feng}},
  \bibinfo{author}{\bibfnamefont{D.}~\bibnamefont{Xiao}},
  \bibinfo{author}{\bibfnamefont{J.}~\bibnamefont{Ding}}, \bibnamefont{and}
  \bibinfo{author}{\bibfnamefont{Y.}~\bibnamefont{Yao}},
  \bibinfo{journal}{Phys. Rev. Lett.} \textbf{\bibinfo{volume}{106}},
  \bibinfo{pages}{016402} (\bibinfo{year}{2011}).

\bibitem[{\citenamefont{Wang et~al.}(2018)\citenamefont{Wang, Wieder, Li, Yan,
  and Bernevig}}]{wangprl2019}
\bibinfo{author}{\bibfnamefont{Z.}~\bibnamefont{Wang}},
  \bibinfo{author}{\bibfnamefont{B.~J.} \bibnamefont{Wieder}},
  \bibinfo{author}{\bibfnamefont{J.}~\bibnamefont{Li}},
  \bibinfo{author}{\bibfnamefont{B.}~\bibnamefont{Yan}}, \bibnamefont{and}
  \bibinfo{author}{\bibfnamefont{B.~A.} \bibnamefont{Bernevig}},
  \bibinfo{journal}{Physical Review Letters} \textbf{\bibinfo{volume}{123}},
  \bibinfo{pages}{186401} (\bibinfo{year}{2018}).

\bibitem[{\citenamefont{Annamamedov et~al.}(1967)\citenamefont{Annamamedov,
  Berger, Petrov, and Slobodchikov}}]{Annamamedov1967}
\bibinfo{author}{\bibfnamefont{R.}~\bibnamefont{Annamamedov}},
  \bibinfo{author}{\bibfnamefont{L.}~\bibnamefont{Berger}},
  \bibinfo{author}{\bibfnamefont{V.}~\bibnamefont{Petrov}}, \bibnamefont{and}
  \bibinfo{author}{\bibfnamefont{S.}~\bibnamefont{Slobodchikov}},
  \bibinfo{journal}{Inorganic Materials} \textbf{\bibinfo{volume}{3}},
  \bibinfo{pages}{1195} (\bibinfo{year}{1967}).

\bibitem[{\citenamefont{Guen et~al.}(1979)\citenamefont{Guen, Glaunsinger, and
  Wold}}]{Guen1979}
\bibinfo{author}{\bibfnamefont{L.}~\bibnamefont{Guen}},
  \bibinfo{author}{\bibfnamefont{W.}~\bibnamefont{Glaunsinger}},
  \bibnamefont{and} \bibinfo{author}{\bibfnamefont{A.}~\bibnamefont{Wold}},
  \bibinfo{journal}{Materials Research Bulletin} \textbf{\bibinfo{volume}{14}},
  \bibinfo{pages}{463} (\bibinfo{year}{1979}).

\bibitem[{\citenamefont{Haeuseler et~al.}(1991)\citenamefont{Haeuseler,
  Ohrendorf, and Himmrich}}]{Haeuseler1991}
\bibinfo{author}{\bibfnamefont{H.}~\bibnamefont{Haeuseler}},
  \bibinfo{author}{\bibfnamefont{F.}~\bibnamefont{Ohrendorf}},
  \bibnamefont{and} \bibinfo{author}{\bibfnamefont{M.}~\bibnamefont{Himmrich}},
  \bibinfo{journal}{Zeitschrift fuer Naturforschung, Teil B. Anorganische
  Chemie, Organische Chemie (33,1978-41,1986)} \textbf{\bibinfo{volume}{46}},
  \bibinfo{pages}{1049} (\bibinfo{year}{1991}).

\bibitem[{\citenamefont{Bradlyn et~al.}(2017)\citenamefont{Bradlyn, Elcoro,
  Cano, Vergniory, Wang, Felser, Aroyo, and Bernevig}}]{tqc2017}
\bibinfo{author}{\bibfnamefont{B.}~\bibnamefont{Bradlyn}},
  \bibinfo{author}{\bibfnamefont{L.}~\bibnamefont{Elcoro}},
  \bibinfo{author}{\bibfnamefont{J.}~\bibnamefont{Cano}},
  \bibinfo{author}{\bibfnamefont{M.~G.} \bibnamefont{Vergniory}},
  \bibinfo{author}{\bibfnamefont{Z.}~\bibnamefont{Wang}},
  \bibinfo{author}{\bibfnamefont{C.}~\bibnamefont{Felser}},
  \bibinfo{author}{\bibfnamefont{M.~I.} \bibnamefont{Aroyo}}, \bibnamefont{and}
  \bibinfo{author}{\bibfnamefont{B.~A.} \bibnamefont{Bernevig}},
  \bibinfo{journal}{Nature} \textbf{\bibinfo{volume}{547}},
  \bibinfo{pages}{298} (\bibinfo{year}{2017}).

\bibitem[{\citenamefont{Hahn and Schulz}(1965)}]{Hahn1965}
\bibinfo{author}{\bibfnamefont{H.}~\bibnamefont{Hahn}} \bibnamefont{and}
  \bibinfo{author}{\bibfnamefont{H.}~\bibnamefont{Schulz}},
  \bibinfo{journal}{Naturwissenschaften} \textbf{\bibinfo{volume}{52}},
  \bibinfo{pages}{426} (\bibinfo{year}{1965}).

\end{thebibliography}

\ \\

\clearpage

\begin{widetext}
\beginsupplement{}
\section*{SUPPLEMENTARY MATERIAL}
\subsection*{A. Topological materials with $\eta=1$}
For the series of the compounds in space group 121, we have systematically computed the band structures and the time-reversal $\mathbb Z_2$ invariants in the two planes: $k_z=0$ and $k_z=\frac{\pi}{c}$. The experimental parameters are employed as reported in the ICSD [shown in Table~\ref{tab:s1}]. We present the band structures of the topological compounds with $\eta=1$. For all these topological compounds, $\nu_{k_z=0}=1$ and $\nu_{k_z=\frac{\pi}{c}}=0$ are in the two planes, respectively. These topological compounds are previously predicted to be TIs~\cite{Wang2010}. However, after we label the irreps of the low-energy bands, one can easily find that they can actually be classified into two cases: \ti~has the $\Gamma_7$ band as the LCB (the upper panels of Fig.~\ref{fig:s1}), while \wsm~has the $\Gamma_6$ band as the LCB (the lower panels of Fig.~\ref{fig:s1}). In the following discussion, we show that the two cases actually correspond to different values of the $S_4~z_2$ indicator.

\begin{figure}[h]
\centering
\includegraphics[height=5.cm]{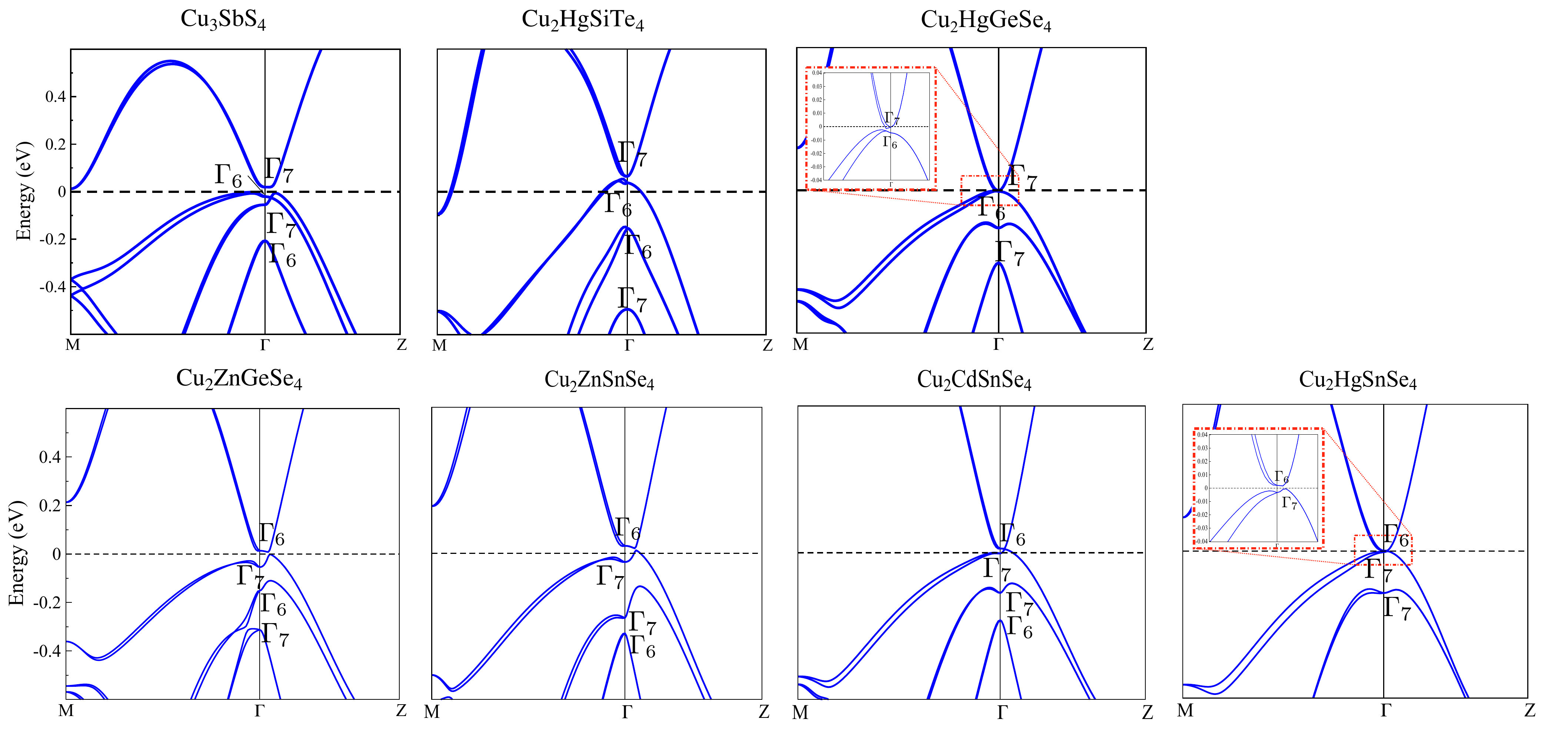}
\caption{(Color online)
The electronic energy bands with SOC and band irreps at the $\Gamma$ point of topological compounds with $\eta=1$. Explicitly, these compounds have a nontrivial $\mathbb Z_2$ invariant in the $k_z=0$ plane, but a trivial $\mathbb Z_2$ invariant in the $k_z=\frac{\pi}{c}$ plane.
}
\label{fig:s1}
\end{figure}

\begin{figure}[!htb]
\centering
\includegraphics[height=3.2cm]{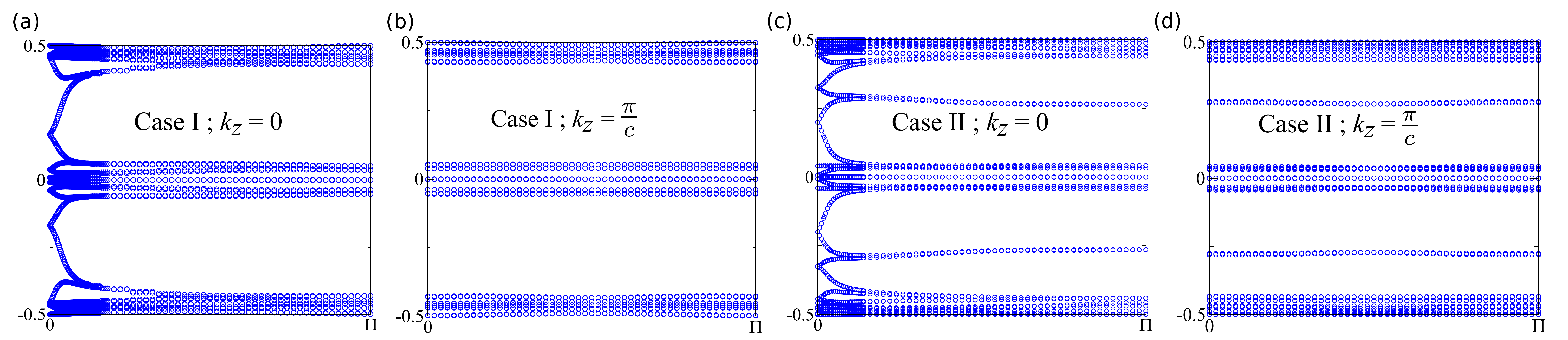}
\caption{(Color online)
The calculated time-reversal $\mathbb Z_2$ in  $k_z=0$ and $k_z={\frac{\pi}{c}}$ planes for \ti~ (a, b) and \wsm~ (c, d), respectively.
} \label{fig:s2}
\end{figure}

\begin{table}[!hb]
\caption{
 ICSD numbers and topological classifications for these topological compounds with SOC.
}\label{tab:s1}
\begin{tabular}{c|c|c|cp{0.5cm}c|c|c|c}
\cline{1-4}\cline{6-9}
\cline{1-4}\cline{6-9}
\cline{1-4}\cline{6-9}
  Compound &ICSD Num. &Previous& This  & &Compound &ICSD Num. & Previous& This work [units: ($\frac{2\pi}{a},\frac{2\pi}{a},\frac{2\pi}{c}$)] \\
   && work& work & & & &  work& \\
\cline{1-4}\cline{6-9}
Cu$_3$SbS$_4$ &  628824\cite{Annamamedov1967}& TI\cite{Wang2010} &TI    && Cu$_2$ZnGeSe$_4$ &627831\cite{Guen1979}&TI\cite{Wang2010}& WSM (0.0036, 0.0, 0.0657)\\
Cu$_2$HgSiTe$_4$ &656152\cite{Haeuseler1991} &TI\cite{tqc2017,nature_2019}&TI && Cu$_2$ZnSnSe$_4$ &629099\cite{Hahn1965}&TI\cite{Wang2010}& WSM (0.0037, 0.0, 0.0757) \\
 Cu$_2$HgGeSe$_4$ & 627692\cite{Hahn1965}&TI\cite{Wang2010}&TI && Cu$_2$CdSnSe$_4$ &619784\cite{Hahn1965}& TI\cite{Wang2010}&WSM (0.0014, 0.0,  0.0294)\\
Cu$_2$HgGeTe$_4$ & 656155\cite{Haeuseler1991}&TI\cite{tqc2017,nature_2019}&TI  && Cu$_2$HgSnSe$_4$ &627936\cite{Hahn1965}&TI\cite{Wang2010}& WSM (0.0049,  0.0,  0.0238)\\
Cu$_3$SbSe$_4$ &  628997\cite{Annamamedov1967}& TI\cite{tqc2017,nature_2019}&TI  && Cu$_2$HgSnTe$_4$ &627940\cite{Haeuseler1991}&Trivial\cite{tqc2017,nature_2019}& WSM (0.0082,  0.0,  0.0338)\\
\cline{1-4}\cline{6-9}
\cline{1-4}\cline{6-9}
\cline{1-4}\cline{6-9}
\end{tabular}
\end{table}

\subsection*{B. The accurate calculations with the mBJ modifications}
Because the band gap is usually underestimated by the PBE functional, we have performed the more accurate calculations by using a modified Beche-Johnson (mBJ) potential. Since the key feature of these band structures is the energy ordering at the $\Gamma$ point, we have systematically presented the evolutions of the energy bands at $\Gamma$ as a function of the mBJ parameter (C$_{\text{mBJ}}$) for different compounds in Fig.~\ref{fig:mbj}.  We found that the relative energies of valence bands almost don't change as varying C$_{\text{mBJ}}$. The energy ordering of \ti~is $\Gamma_7$, $\Gamma_6$ and $\Gamma_7$ (from higher energy to lower energy), while it is $\Gamma_6$, $\Gamma_7$ and $\Gamma_7$ for \wsm. As decreasing the parameter C$_{\text{mBJ}}$, one can clearly see that the energy of the $\Gamma_6$ conduction band decreases monotonically. Accordingly, the $\Gamma_6$ conduction band `intersects' with those three valence bands (guided by eye, in principle, two $\Gamma_6$ bands can not meet each other at $\Gamma$). In Fig.~\ref{fig:mbj}, the critical C$_{\text{mBJ}}$ is denoted by the dashed lines, indicating the band inversion between the $\Gamma_6$ conduction band and the highest valence band.
The results show that the critical C$_{\text{mBJ}}$ parameters for Cu$_2$HgGeTe$_4$, Cu$_3$SbSe$_4$, Cu$_2$HgSnSe$_4$, and   Cu$_2$HgSnTe$_4$ are relatively large, about 1.2, which indicates that the band inversion is more reliable in these compounds.
\begin{figure}[h]
\centering
\includegraphics[height=9cm]{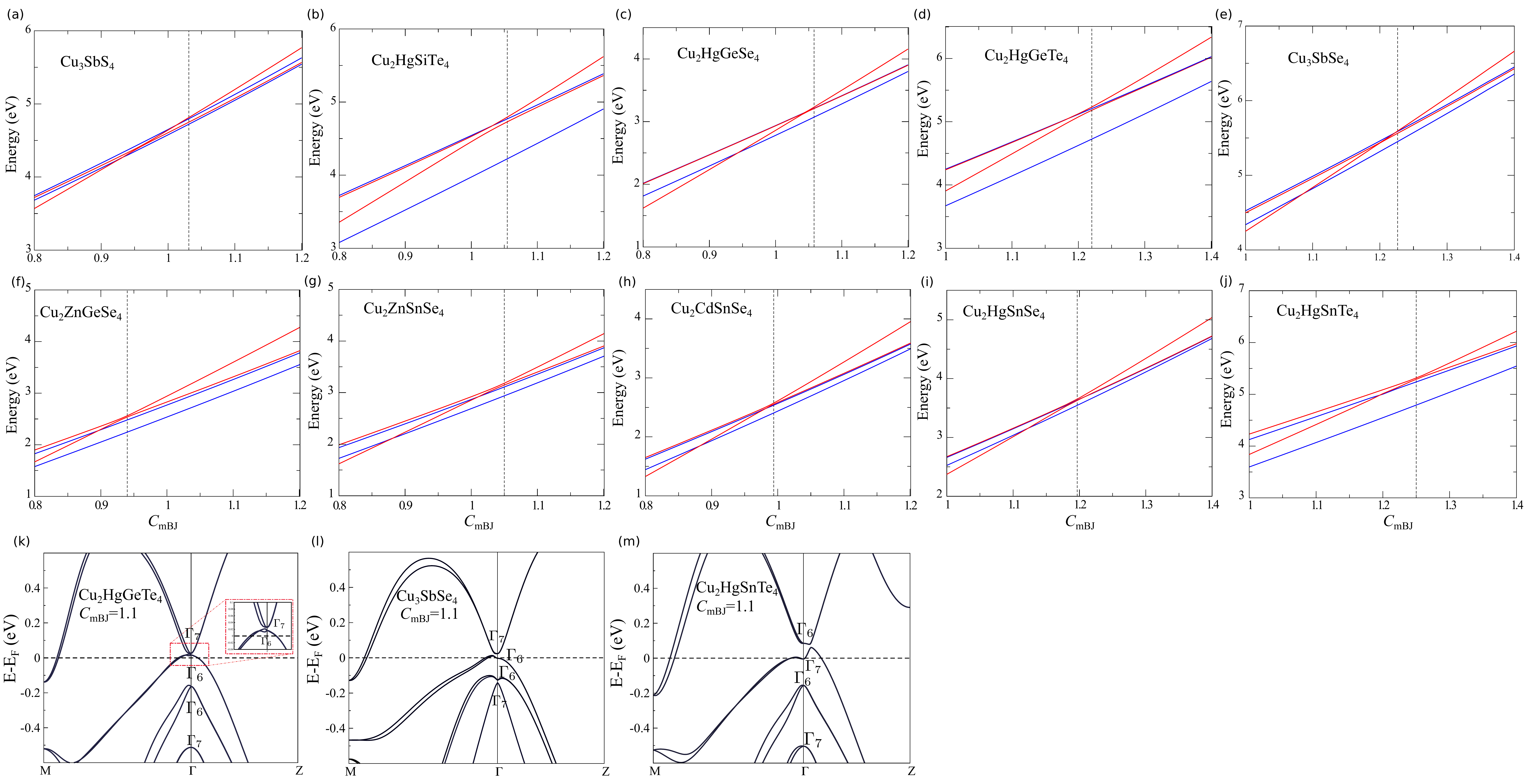}
\caption{(Color online)
(a-j) show the band evolutions of four bands near the Fermi level (three valence bands and one conduction band) at the $\Gamma$ point with varying the parameter $C_\text{mBJ}$ for different topological compounds.  The $\Gamma_6$ and $\Gamma_7$ bands are denoted by the red and blue lines, respectively.
(k-m) present the SOC electronic structures with $C_\text{mBJ}=1.1$ for Cu$_2$HgGeTe$_4$ (k), Cu$_3$SbSe$_4$ (l), and Cu$_2$HgSnTe$_4$ (m), respectively.
}
\label{fig:mbj}
\end{figure}

\subsection*{C. $S_4~z_2$ indicator in these topological compounds}

Since $S_4^4=-1$ in a spinful system, the eigenvalues of $S_4$ symmetry are given as $\lambda_j=e^{i\pi\frac{2j-1}{4}}$ with $j\in\{0,1,2,3\}$. In the TRI systems with $S_4$ symmetry, we propose a generalized definition of the $z_2$ indicator of $S_4$ symmetry:
\begin{equation}
  z_2 = \sum_{K\in\{\text{four SIM}\}}\frac{n_K^{2} - n_K^{0}}{2} \quad{\rm mod} \ 2 ,
\end{equation}
with $n_K^{i}$ the number of the occupied bands with $S_4$ eigenvalue $\lambda_i$  at the SIM $K$.
It will be identical to the definition in Ref.~\cite{song2017,Haruki2018} if all the four SIM are also time-reversal invariant momenta (TRIM).

\begin{figure}[!htb]
\centering
\includegraphics[width=2.5 cm]{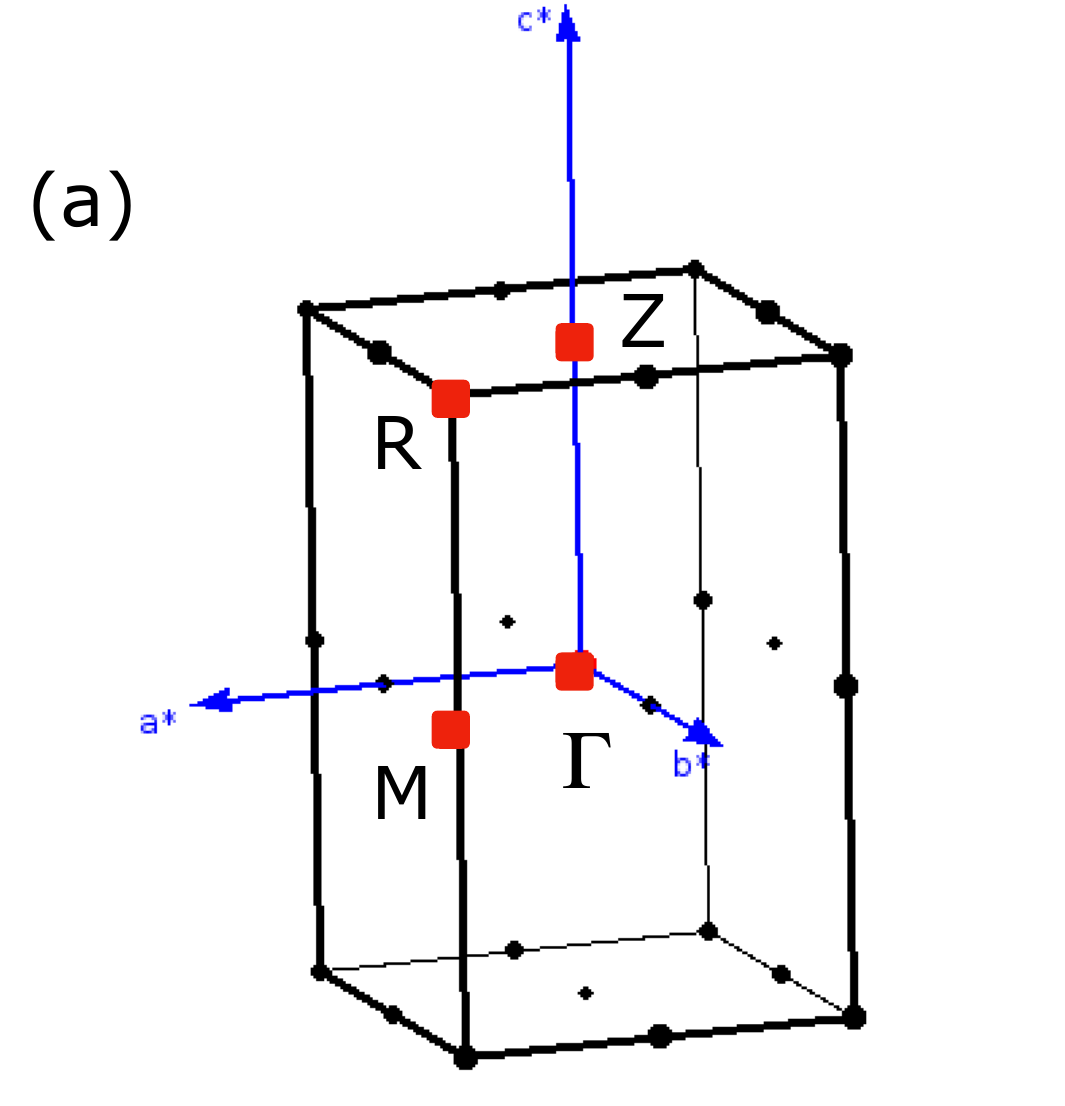}
\includegraphics[width=2.5 cm]{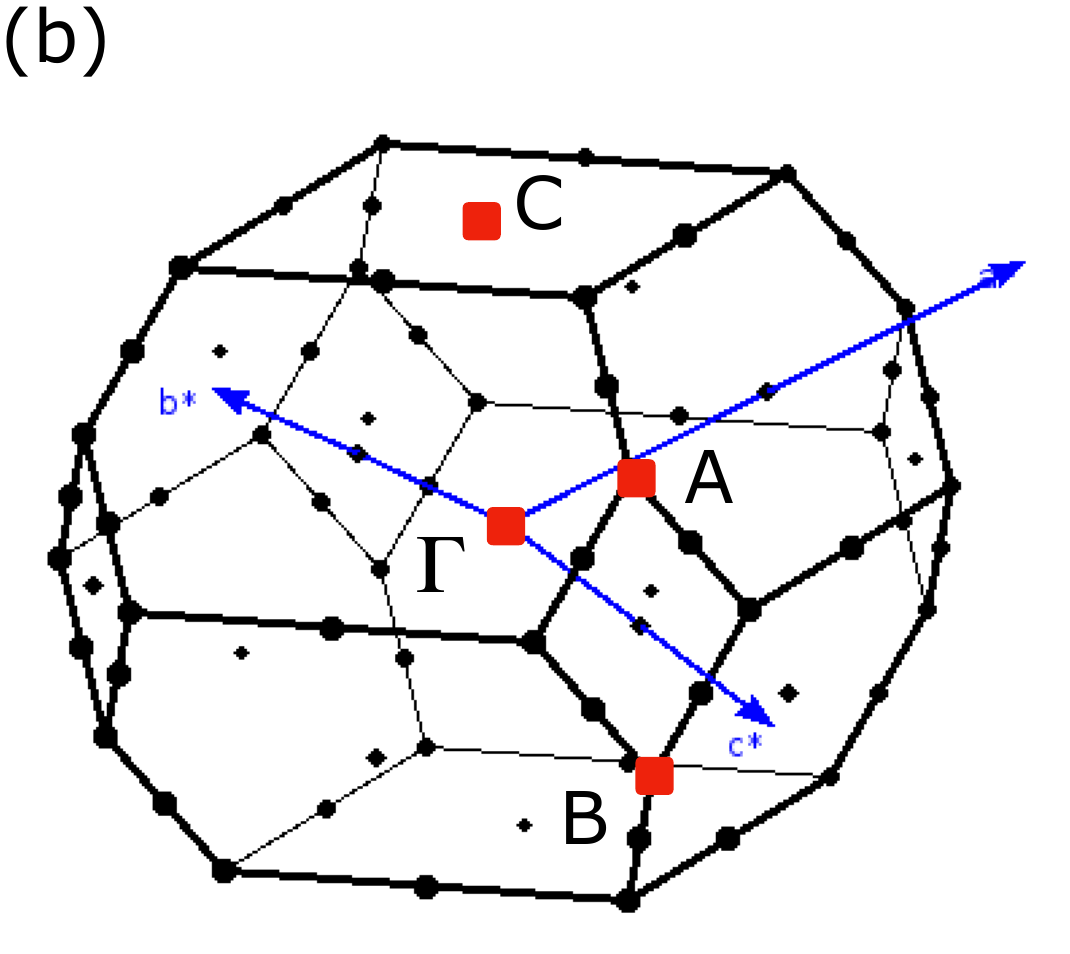}
\includegraphics[width=2.5 cm]{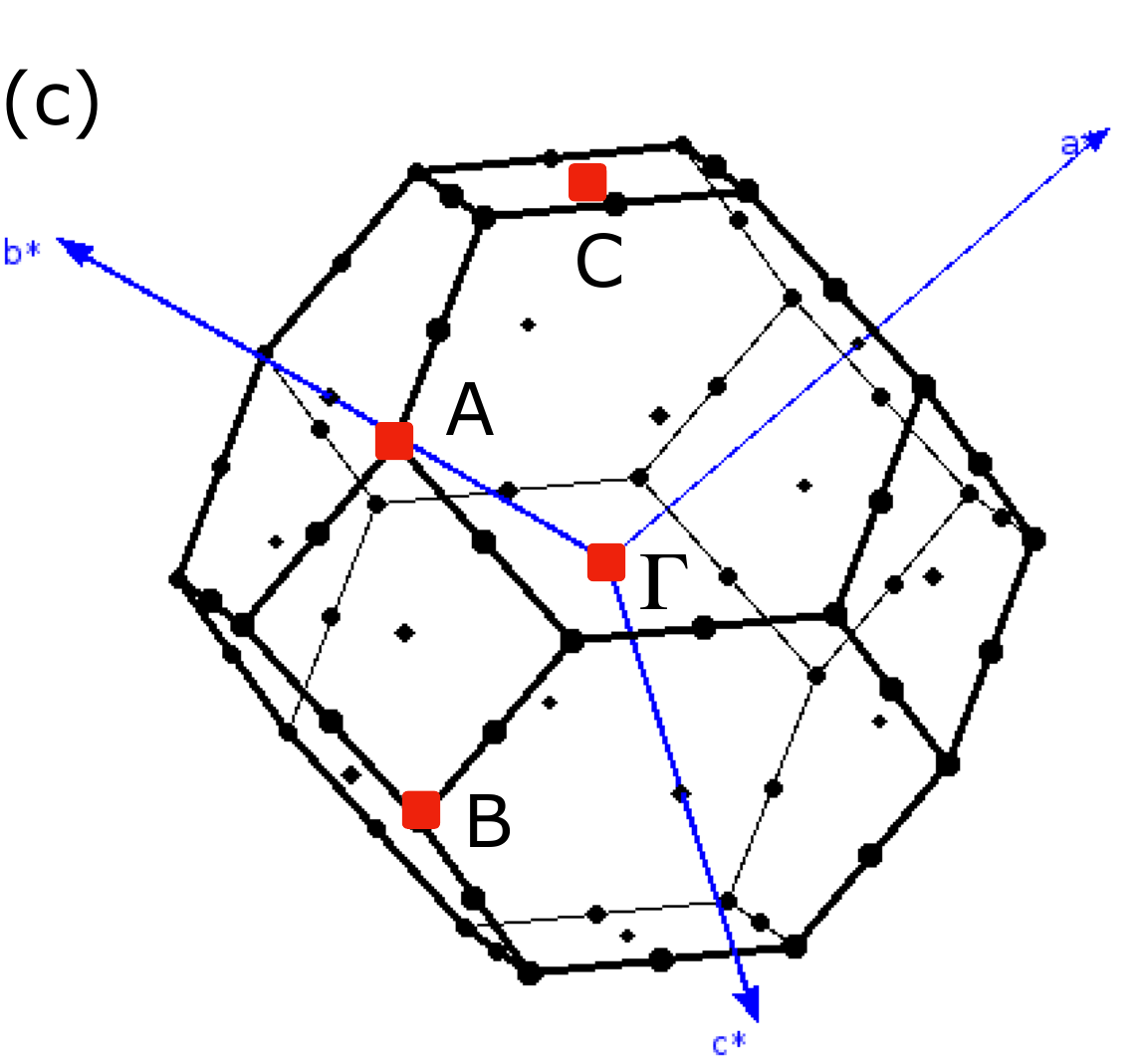}
\caption{(Color online)
The 3D BZs and SIM points. We present the 3D BZs for simple lattice (a), body-centered lattice (b) and face-centered lattice(c), respectively. The SIM points are labeled too.
}
\label{fig:bz}
\end{figure}

In a simple tetragonal structure (SG 81 and its father space groups), the four SIM are $\Gamma[0,0,0]$, $Z[0,0,0.5]$, $M[0.5,0.5,0.0]$, and $R[0.5,0.5,0.5]$ (The $k$ points are given in units of [$\frac{2\pi}{a},\frac{2\pi}{a},\frac{2\pi}{c}$] in Cartesian coordinates). Since the four SIM are also TRIM, all the bands are doubly-degenerate. Thus, $n_K^{\frac{1}{2}}~(n_K^{\frac{3}{2}})$ equals to $n_K^{0}~(n_K^{2})$.
As defined in Ref.~\cite{song2017,Haruki2018}, $n_K^{\frac{1}{2}}~(n_K^{\frac{3}{2}})$ is the number of Kramers pairs at $K$ with tr$[D(S_4)]=\sqrt{2}$ (with tr$[D(S_4)]=-\sqrt{2}$), and $D(S_4)$ is the representation matrix on the corresponding Kramers pair.
\begin{equation}
  z_2 = \sum_{K\in\{\Gamma,Z,M,R\}}\frac{n_K^{\frac{3}{2}} - n_K^{\frac{1}{2}}}{2} \quad{\rm mod} \ 2 ,
\end{equation}

In a body-centered tetragonal structure (SG 82 and its father space groups), four SIM are $\Gamma[0,0,0]$, $C[0,0,1]$, $A[0.5,0.5,0.5]$, and $B[0.5,0.5,-0.5]$ (The $k$ points are given in units of [$\frac{2\pi}{a},\frac{2\pi}{a},\frac{2\pi}{c}$] in Cartesian coordinates). Note that the $A$ and $B$ points are not TRIM. Namely, there is no Kramers pair at $A$ and $B$. So, the $S_4$ $z_2$ indicator is redefined as,
\begin{equation}
  z_2 = \sum_{K\in\{\Gamma,C,A,B\}}\frac{n_K^{2} - n_K^{0}}{2} \quad{\rm mod} \ 2 ,
\end{equation}
Since space group 121 has a body-centered tetragonal structure, $n_K^{0}$ and $n_K^{2}$ are computed for four SIM and listed explicitly in Table~\ref{tab:weyls4}.
The $z_{2}$ indicator is computed to be 1 and 0 for \ti~and \wsm, respectively.

\begin{table}[!h]
\caption{
The number of occupied Kramers pairs with $S_4$ eigenvalue $\lambda_2$ and $\lambda_0$ on SIM. The last column shows the $S_4$ $z_2$ indicator calculated by using these symmetry data.
  }\label{tab:weyls4}
  \begin{tabular}{c|c|c|c|c|c}
  \hline
  \hline
  $n_K^{2},n_K^{0}$ & $\Gamma$ & C & A & B & $S_4~z_2$  \\
  \hline
   \ti~(TI)   & 15,16 & 16,15 & 16,15 & 16,15 & 1 \\
   \wsm~(WSM) & 14,17 & 16,15 & 15,16 & 15,16 & 0 \\
  \hline
  \hline
  \end{tabular}
\end{table}

In a face-centered cubic structure (SG 216 and its father space groups), four SIM are $\Gamma[0,0,0]$, $C[0,0,1]$, $A[1,0,0.5]$, and $B[1,0,-0.5]$ (the $k$ points are given in units of [$\frac{2\pi}{a},\frac{2\pi}{a},\frac{2\pi}{a}$] in Cartesian coordinates). Note that $A$ and $B$ points are not TRIM. Namely, there is no Kramers pair at $A$ and $B$. So, the $S_4$ $z_2$ indicator is redefined as,
\begin{equation}
  z_2 = \sum_{K\in\{\Gamma,C,A,B\}}\frac{n_K^{2} - n_K^{0}}{2} \quad{\rm mod} \ 2 ,
\end{equation}

\clearpage
\subsection*{D. Six-band model}
Using the $\Gamma_7^-$ and $\Gamma_8^+$ bands under the symmetry of $O_h$, one can construct a six-band effective model. Explicitly, under the basis of $\{i|xyz\uparrow\rangle, i|xyz\downarrow\rangle, |\frac{3}{2}, \frac{3}{2}\rangle, |\frac{3}{2}, \frac{1}{2}\rangle, |\frac{3}{2}, -\frac{1}{2}\rangle, |\frac{3}{2}, -\frac{3}{2}\rangle\}$, the $O_h$-invariant $\bold{k\cdot p}$ Hamiltonian can be given as follows:
\begin{equation}
    \begin{split}
        &H'= \begin{bmatrix}
            \left(A_0+A_2k^2\right) {\mathbb I}_{2}& C_3{\mathbb S}^\dagger \\
            C_3{\mathbb S} & H_0
        \end{bmatrix}\text{ with}~H_0=(B_0+B_2k^2){\mathbb I}_{4}+C_1 {\mathbb E}+C_2{\mathbb T}, \\
 \text{where}  &~  k\equiv k_x^2+k_y^2+k_z^2 \text{~and ${\mathbb I}_n$ is an $n$-dimensional identity matrix,} \\
        &{\mathbb E} = \begin{pmatrix}
            2k_z^2-k_x^2-k_y^2 & 0 & \sqrt{3}(k_x^2-k_y^2) & 0 \\
            0 & -(2k_z^2-k_x^2-k_y^2) & 0 & \sqrt{3}(k_x^2-k_y^2) \\
            \sqrt{3}(k_x^2-k_y^2) & 0 & -(2k_z^2-k_x^2-k_y^2) & 0 \\
            0 & \sqrt{3}(k_x^2-k_y^2) & 0 & 2k_z^2-k_x^2-k_y^2
        \end{pmatrix}\\
        &{\mathbb T} = \begin{pmatrix}
            0 & k_- k_z & -ik_x k_y & 0 \\
            k_+ k_z & 0 & 0 & -ik_x k_y \\
            ik_x k_y & 0 & 0 & -k_- k_z \\
            0 & i k_x k_y & -k_+ k_z & 0
        \end{pmatrix},~
    {\mathbb S} = \begin{pmatrix}
        k_+ & 2k_z \\
        0 & -\sqrt{3}k_+ \\
        \sqrt{3}k_- & 0 \\
        2k_z & -k_-
    \end{pmatrix}
    \end{split}
\end{equation}
Also, the matrix representations of the generators of $O_h$ are given in Table~\ref{tab:rep}.
\begin{table}[!h]
  \caption{
The matrix representations of the generators (\ie $C_{3,111}$ and $C_{4z}$) of $O_h$, given under the basis of $\Gamma_7^-$ and $\Gamma_8^+$, respectively.
  }\label{tab:rep}
  \begin{tabular}{c|c|c}
  \hline
  \hline
                &     $\Gamma_7^-$   &   $\Gamma_8^+$ \\
  \hline
   C$_{3,111}$  &     $\frac{1}{2}\left(\begin{array}{cc} 1-i & -1-i \\ 1-i & 1+i \end{array}\right)$   &   $\frac{1}{4}\left(\begin{array}{cccc} -1-i & -\sqrt{3}+\sqrt{3}i & \sqrt{3}+\sqrt{3}i & 1-i \\ -\sqrt{3}-\sqrt{3}i & -1+i & -1-i & -\sqrt{3}+\sqrt{3}i \\ -\sqrt{3}-\sqrt{3}i & 1-i & -1-i & \sqrt{3}-\sqrt{3}i \\ -1-i & \sqrt{3}-\sqrt{3}i & \sqrt{3}+\sqrt{3}i & -1+i \end{array}\right)$ \\
  \hline
   C$_{4z}$     &    $-\frac{\sqrt{2}}{2}\left(\begin{array}{cc} 1-i & 0 \\ 0 & 1+i \end{array}\right)$   &   $\left(\begin{array}{cccc} -(-1)^{\frac{1}{4}} & 0 & 0 & 0 \\ 0 & -(-1)^{\frac{3}{4}} & 0 & 0 \\ 0 & 0 & (-1)^{\frac{1}{4}} & 0 \\ 0 & 0 & 0 & (-1)^{\frac{3}{4}} \end{array}\right)$ \\
  \hline
   $I$ & $-{\mathbb I_2}$&  ${\mathbb I_4}$\\
  \hline
   $\cal{T}$     &    $-\left(\begin{array}{cc} 0 & -1 \\ 1 & 0 \end{array}\right)K $ & $ \left(\begin{array}{cccc}0 & 0 & 0 & 1 \\ 0 & 0 & -1 & 0 \\ 0 & 1 & 0 & 0 \\ -1 & 0 & 0 & 0\end{array}\right)K $ \\
  \hline
  \hline
  \end{tabular}
\end{table}

To obtain the $D_{4h}$ symmetry, one can easily change $A_2k^2~(B_2k^2)$ to $A_1k_z^2+A_2k_{||}^2~(B_1k_z^2+B_2k_{||}^2)$ and add another diagonal term $H_A$ to $H_0$, which is a uni-axial strain in the $z$-axis. Simply, $H_A$ can take the form of $Diag\{1,-1,-1,1\}$.
Then, in order to break $I$ and $C_{4z}$ but keep S$_{4z}$, $H_B$ (first-order of $\bk$) and $H_C$ (first-order of $\bk$) are added. The $D_{2d}$-invariant Hamiltonian is derived as
\begin{equation}
    \begin{split}
        &H(\bk) = \begin{bmatrix}
            \left(A_0+A_1k_z^2+A_2k_{||}^2\right) {\mathbb I}_{2}& C_3{\mathbb S}^\dagger \\
            C_3{\mathbb S} & \left(B_0+B_1k_z^2+B_2k_{||}^2\right){\mathbb I}_{4}+C_1 {\mathbb E}+C_2{\mathbb T}+\delta_1 H_A+\delta_2 H_B +\delta_3 H_C
        \end{bmatrix}
    \end{split}
\end{equation}
with additional first-order term $H_B$,
\begin{equation}
  H_B = \begin{pmatrix}
      0 & -k_+ & 2k_z & -\sqrt{3}k_- \\
      -k_- & 0 & \sqrt{3}k_+ & -2k_z \\
      2k_z & \sqrt{3}k_- & 0 & -k_+ \\
      -\sqrt{3}k_+ & -2k_z & -k_- & 0
  \end{pmatrix}
\end{equation}
and the third-order term $H_C$,
\begin{equation}
    H_{C}= k_z(k_x^2-k_y^2)J_z+ k_x(k_y^2-k_z^2)J_x+ k_y(k_z^2-k_x^2)J_y
\end{equation}
with
\begin{equation}
    J_x=\begin{pmatrix}
       0 & \sqrt{3} & 0 & 0\\
       \sqrt{3} & 0 & 2 & 0\\
       0 & 2 & 0 & \sqrt{3}\\
       0 & 0 & \sqrt{3} & 0
    \end{pmatrix};~
    J_y=\begin{pmatrix}
       0 & -\sqrt{3}i & 0 & 0\\
       \sqrt{3}i & 0 & -2i & 0\\
       0 & 2i & 0 & -\sqrt{3}i\\
       0 & 0 & \sqrt{3}i & 0
    \end{pmatrix};~
    J_z=\begin{pmatrix}
       3 & 0 & 0 & 0\\
       0 & 1 & 0 & 0\\
       0 & 0 &-1 & 0\\
       0 & 0 & 0 &-3
    \end{pmatrix};~
\end{equation}
With the parameters given in the main text, the band structures of the model are obtained in Fig.~\ref{fig:s3}.
\begin{figure}[h]
\centering
\includegraphics[height=5cm]{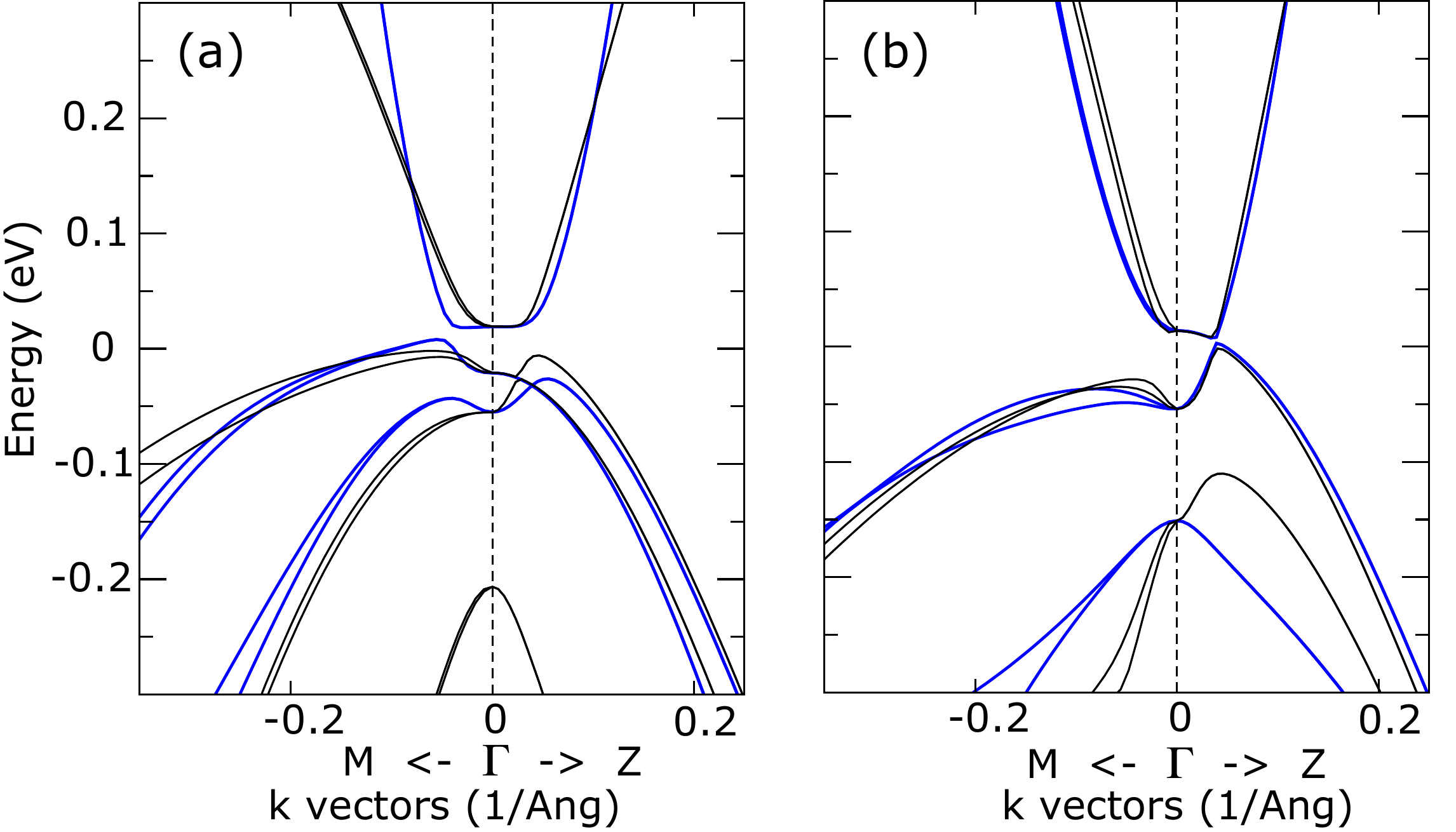}
\caption{(Color online)
The fitted electronic energy bands around the $\Gamma$ point of ~\tic~(a) and ~\wsmc~(b) with fitting parameters shown in Table~\ref{tab:matpara}, where the black lines are the band structures from first-principles calculations and the blue lines are the results of the fitted effective six-band model.
}
\label{fig:s3}
\end{figure}

\clearpage

\end{widetext}
\end{document}